\documentclass[twocolumn,a4paper,english,preprintnumbers,amsmath,amssymb]{revtex4}
\usepackage[latin1]{inputenc}
\usepackage{babel}
\usepackage{array}
\usepackage{graphicx}
\usepackage{color}
\usepackage{listings}

\makeatletter

\begin{document}

\title{A parameter-free, solid-angle based, nearest-neighbor algorithm}

\author{Jacobus A. van Meel}
\affiliation{FOM Institute for Atomic and Molecular Physics, Science Park 104, 1098 XG Amsterdam, The Netherlands}

\author{Laura Filion}
\affiliation{Department of Chemistry, University of Cambridge, Lensfield Road, Cambridge, CB2 1EW, United Kingdom}

\author{Chantal Valeriani}
\affiliation{SUPA, School of Physics and Astronomy, University of Edinburgh, Mayfield Road, Edinburgh, EH9 3JZ, Scotland}
\affiliation{Departamento de Quimica Fisica, Facultad de Quimica, Universidad Complutense de Madrid, 28040, Madrid, Spain}

\author{Daan Frenkel}
\affiliation{Department of Chemistry, University of Cambridge, Lensfield Road, Cambridge, CB2 1EW, United Kingdom~}

\begin{abstract}
We propose a parameter-free algorithm for the identification of nearest neighbors. 
The algorithm is very easy to use and has a number of advantages over existing  algorithms to identify nearest-neighbors.
This solid-angle based nearest-neighbor algorithm (SANN) attributes to each possible neighbor a solid angle 
and determines the cutoff radius by the requirement that the sum of the solid angles is $4\pi$.  
The algorithm can be used to analyze 3D images, both from experiments as well as theory, and as the  algorithm  
has a low computational cost, it can also be used ``on the fly'' in simulations.
In this paper, we describe the SANN algorithm, discuss
its properties, and compare it to both a fixed-distance cutoff algorithm and to a Voronoi construction by
analyzing its behavior in bulk phases of systems of carbon atoms, Lennard-Jones particles and hard spheres
as well as in Lennard-Jones systems with liquid-crystal and liquid-vapor interfaces.
\end{abstract}


\maketitle

\section{Introduction}

In most studies of many-particle systems, one is confronted with the task of
determining the nearest neighbors of a particle, or set of particles.
Interestingly, while identifying nearest neighbors is an important component of various analyses,
and is sometimes even needed to evaluate interaction potentials, there is no unique definition of a
nearest neighbor and, as a result, what one defines as a nearest neighbor
is typically dependent on the question at hand. The two most common algorithms
for determining nearest neighbors are i) a fixed-distance cutoff and ii) a
Voronoi construction~\cite{Okabe_2000}.  However, many extensions, and other 
definitions have been used as well, see for instance Refs.~\cite{Knuth_1968,randomalgo}.

A fixed-distance cutoff is the obvious choice in simulations with particles
interacting through a short-range potential, where each nearest neighbor is an interaction
partner and the cutoff distance corresponds to the interaction range. Additionally,
fixed-distance cutoffs have also been used in determining nearest neighbors for structural analyses
 such as calculating bond-order parameters in nucleation studies (e.g. see Ref.~\cite{Wolde_1996a}).
However, in these cases, the ``fixed-distance'' is not well defined.  Arguably, the
first minimum of the pair correlation function $g(r)$ (also known as radial distribution
function) is a reasonable choice for
the cutoff, as it relates to the neighbors in the first coordination shell.
However, the precise location of this minimum depends on both the system's details
and thermodynamic conditions and therefore must be determined every time either one is
changed. Additionally, the cutoff is defined for the entire system and, as such, is
not appropriate for systems with large density gradients, such as
occur naturally in nucleation studies, in systems in the presence of gravity 
or in systems with interfaces.

In contrast, a Voronoi construction~\cite{Okabe_2000} is based on purely
geometric constraints and is parameter free. In addition to identifying
nearest neighbors, this method can be used to determine geometric properties like
edges and faces shared between these neighbors - data that are frequently useful
for structural analysis and classification. Based on the local environment around a
particle, it is more appropriate than a fixed-distance cutoff in the case of density
gradients. However, there are also a number of inherent problems
with a Voronoi construction, some of which will be discussed in this
manuscript. First of all, the method is computationally expensive and hence is rarely used
on-the-fly in simulations. More importantly, it is not robust  against
thermal fluctuations. In a crystal, thermal fluctuations which cause particles to fluctuate
around their equilibrium lattice sites can spuriously increase the number of particles which
share a small face with the target particle~\cite{Tanemura_1977,Troadec_1998} and hence
increase the number of particles identified as nearest neighbors. There exist extensions to the
Voronoi construction which aim to increase the robustness against these
fluctuations~\cite{Tanemura_1977,Hsu_1979,Brostow_1998,Bandyopadhyay_2004,Yu_2005},
however, they typically introduce parameters, removing the ``parameter free''
advantage of the algorithm, and they further increase the computational cost.


Looking at the advantages and disadvantages of both the fixed-distance cutoff and the
Voronoi algorithm, we suggest a list of features which a ``good'' nearest neighbor
algorithm should have. Specifically, an algorithm should i) be able to deal with systems
with inhomogeneous density, ii) be stable against thermal fluctuations,  iii) be parameter free 
and iv) be computationally inexpensive. In this manuscript, we propose, with these
goals in mind,  a simple algorithm for the identification of nearest neighbors: the
solid-angle based nearest-neighbor algorithm (SANN). This method is based on similar principles to a
theory used by Corwin {\it et al.}~\cite{Corwin_2010} which used solid angles to predict 
the number of nearest neighbors. 
It is also similar to a 
Voronoi construction as it does not require tuneable parameters. However, SANN is 
computationally  significantly less expensive than a Voronoi construction. In fact, its computational cost only slightly
exceeds that of a fixed-distance cutoff  making it suitable for on-the-fly use in
simulations. In order to compare our algorithm with the fixed-distance cutoff and
Voronoi construction, we apply all three methods to monodisperse hard spheres, 
Lennard-Jones liquid and fcc crystal bulk phases, the $3$-fold coordinated liquid
carbon and graphite phases and the $4$-fold coordinated liquid carbon and diamond
phases and compare the set of nearest neighbors obtained with the three methods.
On some liquid/solid systems we also compute  bond-order parameters and discuss the impact 
different nearest-neighbor sets have on the bond-order correlator distributions. 
To conclude, we study liquid-crystal and liquid-vapor Lennard-Jones two-phase systems
to test the behavior of SANN at interfaces.


\section{Method}\label{sec:method}
\subsection{Description of SANN method}
\begin{figure}
\includegraphics[width=0.45\textwidth]{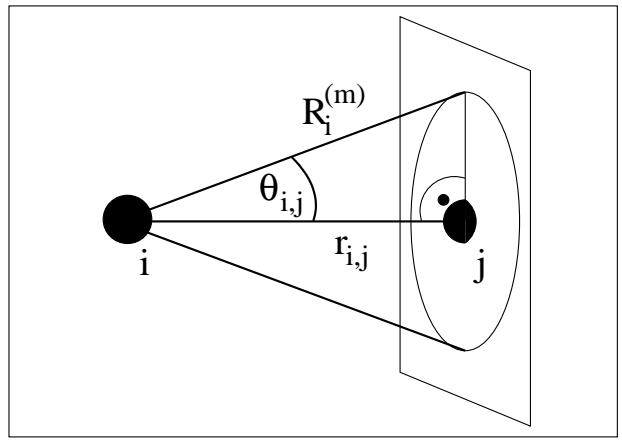}
\caption{Definition of the angle $\theta_{i,j}$ associated with a neighbor $j$
of particle $i$. Here, $r_{i,j}$ is the distance between both particles and $R^{(m)}_i$
is the neighbor shell radius.}
\label{fig:solid_angle}
\end{figure}

As mentioned in the introduction, there exists no unique definition of a nearest neighbor.
Consequently, it comes as no surprise that our SANN algorithm introduces a definition
that differs from those of existing algorithms. Yet it has many similarities with the
definitions of the fixed-distance cutoff and the Voronoi construction, as it is based
on similar concepts.

\begin{figure*}
\center
\includegraphics[width=0.95\textwidth]{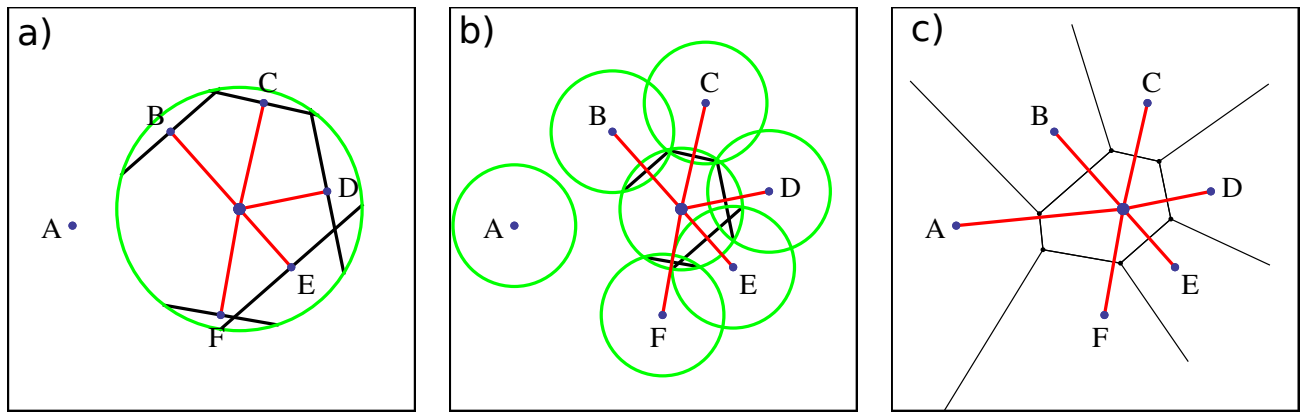}
\caption{$2d$ comparison of the SANN and Voronoi algorithms.  In all three
panels we show a central particle with potential nearest neighbors A through F.  
In panel a) we sketch the SANN algorithm: the green circle shows the shell radius, and particles
B through F are identified as nearest neighbors. Note that in all panels
the nearest neighbors are indicated with red lines. To facilitate 
the comparison between a Voronoi construction and the SANN algorithm, in panel b)  we make use of the 
fact that the SANN algorithm is scale free, i.e. $r_{i,j}/R^{(m)}_i$ = $(0.5 r_{i,j})/(0.5 R^{(m)}_i)$ where  $0.5 r_{i,j}$ 
is simply the distance from particle $i$ to the midpoint between $i$ and $j$, and $0.5 R^{(m)}_i$ 
is half the shell radius. In panel b) the green circles have a radius equal to the half the shell radius of the center particle, 
and are centered around each particle; the black lines are constructed by finding the 
intersection of the green circles and indicate the width of the solid angle between the center
particle and each of its neighbors, respectively.  Finally, in panel c) we show a Voronoi construction for the same set of particles.
Note that the Voronoi construction finds an extra nearest neighbor, i.e. particle A.
}
\label{fig:comparison}
\end{figure*}
Consider a dense system with excluded volume, where we have a particle $i$ located
at position $\vec{r}_i$ surrounded by particles $\{j\}$. The fixed-distance cutoff defines the 
nearest neighbors of particle $i$ to be all the particles of $\{j\}$ with a distance to $i$
smaller than the cutoff-distance.  However, as mentioned in the introduction, the problem
with this definition is in choosing that distance. This is where our SANN algorithm comes
into play. For each particle $i$ SANN determines an \emph{individual}
cutoff distance $R^{(m)}_{i}$, which we call the \emph{shell radius}. It depends 
on the local environment of particle $i$ and includes its $m$ nearest neighbors. Since 
the cutoff distance is now a local property, the algorithm is suitable for systems with
inhomogeneous densities. For the computation of $R^{(m)}_i$ SANN uses a purely geometrical
construction, as does the Voronoi tessellation. Thus, the algorithm is parameter-free and
scale-free. In the following we describe the geometrical construction and how $m$ and
$R^{(m)}_{i}$ are determined. 

First, we assume the particles $\{j\}$ surrounding $i$ are known and ordered 
such that $r_{i,j} \le r_{i,j+1}$ for all $j$. This relates
$R^{(m)}_i$ and $m$ in the following manner:
\begin{equation}
r_{i,m} \le R^{(m)}_i < r_{i,m+1}.
\label{eqn:property}
\end{equation}
Then, starting with the particle closest to $i$ we associate with each potential neighbor $j$
an angle $\theta_{i,j}$ based on the distance between the particles $r_{i,j} = \left| \vec{r}_j - \vec{r}_i \right|$
and the yet undetermined shell radius $R^{(m)}_i$ as depicted in Figure~\ref{fig:solid_angle}.

SANN defines the neighborhood of a particle $i$ to consist of the nearest (i.e. closest)
$m$ particles $\{j\}$ such that the sum of their solid angles associated with $\theta_{i,j}$
equals $4 \pi$, i.e. 
\begin{equation}
4 \pi = \sum_{j=1}^{m} 2 \pi [1 - cos( \theta_{i,j} )] = 
\sum_{j=1}^{m} 2 \pi (1 - r_{i,j} / R^{(m)}_i).
\label{eqn:sann}
\end{equation}
We point out that while the number $m$ and the shell radius $R^{(m)}_i$ are not known yet, they are not
independent: once one is known it is straightforward to determine the other. Also note that since the
solid angle contribution for a single neighbor is always less than $2\pi$, $m$ must be at least 3.

To visualize this idea imagine each solid angle as a cone with its apex point located at particle $i$
and the cone's base center located at neighbor $j$. For a complete set of nearest neighbors all those cones
stack to fill a spherical volume around $i$ with a radius corresponding to the shell radius $R^{(m)}_i$.
Obviously, cones are not space-filling (i.e. they don't stack without gaps), so for the sum of solid
angles to equal $4\pi$ some cone overlap does occur.

Combining Eqn.~\ref{eqn:property} and Eqn.~\ref{eqn:sann} leads to a condition for
the determination of the neighbor shell radius,
\begin{equation}
R^{(m)}_i = \frac{ \sum_{j=1}^{m} r_{i,j} }{ m - 2 } < r_{i,m+1},
\label{eqn:condition}
\end{equation}
where $R^{(m)}_i$ refers to the shell radius containing $m$ particles. To solve this
inequality, we start with the smallest number of neighbors capable of satisfying
Eqn.~\ref{eqn:sann}, $m=3$, and increase $m$ iteratively. During each iteration, we evaluate   
Eqn.~\ref{eqn:condition}  and the smallest $m$ that satisfies the equation
yields the number of neighbors $N_b(i)$ with $R^{(m)}_i$ the corresponding neighbor shell radius.
It is straightforward to show that the algorithm converges, because the neighbor
distance increases monotonically due to the sorting, $r_{i,m+1} \ge r_{i,m}$, and the
cutoff radius $R^{(m)}_i$ decreases monotonically, $R^{(m+1)}_i \le R^{(m)}_i$.

To highlight the differences and similarities between the geometry of the SANN algorithm and 
that of the Voronoi construction, we show in Figure~\ref{fig:comparison} $2d$-schematics of both.
In panel a) we depict the SANN algorithm: the shell radius is show as a green circle, and the red
lines connect the central particle to the nearest neighbors determined by SANN, i.e. particles B
through F. In panel c) we 
show the associated  Voronoi construction. We note that it identifies all particles A to F as neighbors of the
center particle. The Wigner-Seitz cell is indicated with black lines. Neighbor
A shares only a small face and is fragile to thermal fluctuations. 
To compare the Voronoi construction to the SANN algorithm, in panel b) we make use of the
fact that the algorithm is scale free, i.e. $r_{i,j}/R^{(m)}_i$ = $(0.5 r_{i,j})/(0.5 R^{(m)}_i)$ where  $0.5 r_{i,j}$
is simply the distance from particle $i$ to the midpoint between $i$ and $j$, and $0.5 R^{(m)}_i$
is half the shell radius. In panel b) the green circles have a radius equal to the half the shell radius of the center particle,
and are centered around each particle; the black lines are constructed by finding the
intersection of the green circles and indicate the width of the solid angle between the center
particle and each of its neighbors, respectively.
Hence, one way to picture this method is to picture slowly growing spheres around each particle.
When the intersecting planes associated with particle $i$ (the black lines in the plot) yield solid angles
summing to 4$\pi$, the shell radius of particle $i$ has been found. This procedure is then repeated for each particle.
In the schematic (Figure 2), Particle A is
not a neighbor since there is no overlap between the green circle around particle A and green circle around the center particle,
hence the fragility problem highlighted in the discussion of the Voronoi construction is not present here.
The black lines indicate the width of the solid angles and can be compared to the faces of the Wigner-Seitz cell.
However, the faces are not identical to the real Wigner-Seitz cell. In general the faces are either larger
or smaller than the real Wigner-Seitz faces. Note that, by definition, SANN extends each face to the shell
circle (see Fig.~\ref{fig:solid_angle}), hence, the black lines in the SANN algorithm overlap sometimes, i.e. between
particle D and E as well as E and F.

\subsection{Algorithm}\label{sec:scheme}
Following the procedure outlined in the previous section we propose this
simple scheme to determine the nearest neighbors of particle $i$:
\begin{enumerate}
\item Compute distances $r_{i,j}$ to all potential neighbors $\{j\}$ from $i$.
\item Sort possible neighbors $\{j\}$ by their distance $r_{i,j}$ in increasing order.
\item Start with $m = 3$ (i.e. the minimum number of neighbors).
\item Compute $R^{(m)}_i = \sum_{j=1}^{m} r_{i,j} / (m - 2)$.
\item If $(R^{(m)}_i > r_{i,m+1})$, then increment $m$ by $1$ and go back to step 4.
\item Otherwise, $m$ is $N_b(i)$, i.e. the number of neighbors for particle $i$,
      and $R^{(m)}_i$ the associated neighbor shell radius.
\end{enumerate}

A C/Fortran implementation of the scheme can be found in the Supplementary Materials~\cite{SI}.

 \subsection{Algorithm properties} \label{sec:properties}

Before comparing our algorithm to the results from a fixed-distance cutoff and a Voronoi
construction, we first discuss several inherent properties of our SANN algorithm.

\begin{description}

\item[Convergence:]
Provided there exist enough neighbors $\{j\}$ of particle $i$ the algorithm converges,
because $R^{(m+1)}_i < R^{(m)}_i$ and all neighbors are sorted such that $r_{m+1} \ge r_{m}$. To proof
this, we express $R^{(m+1)}_i$ in terms of $R^{(m)}_i$:
\begin{equation}
R^{(m+1)}_i = 
R^{(m)}_i \Big[ \frac{ m-2 + r_{i,m+1} / R^{(m)}_i }{ m-2+1 } \Big].
\label{eqn:convergence}
\end{equation}
The definition of $R^{(m+1)}_i$ requires that $R^{(m)}_i > r_{i,m+1}$, which in combination with
Eqn.~\ref{eqn:convergence} leads to $R^{(m+1)}_i < R^{(m)}_i$.

\item[Equal distances neighbors:]
The algorithm also ensures that multiple neighbors with equal distance to the center particle are
all identified as neighbors. To proof this we show that $R^{(m+1)}_i > r_{i,m+i}$, which means that
the SANN radius is always larger than each particle distance included. Analogous to the convergence
proof, we express $R^{(m+1)}_i$ in terms of $r_{i,m+1}$, the latest (and largest) distance included:
\begin{equation}
R^{(m+1)}_i = 
r_{m+1} \Big[ \frac{ R^{(m)}_i / r_{i,m+1} (m-2) + 1 }{ m - 2 + 1 } \Big].
\label{eqn:degeneration}
\end{equation}
Again, $R^{(m+1)}_i$ requires that $R^{(m)}_i > r_{i,m+1}$, which combined with Eqn.~\ref{eqn:degeneration}
leads to $R^{(m+1)}_i > r_{i,m+1}$. Therefore, if multiple neighbors share the same distance, all are
included.

\item[Pair-wise symmetry:]
For both a fixed-distance cutoff and the Voronoi construction, the neighbors are symmetric in
the sense that if particle $i$ is a neighbor of $j$, then $j$ is also a neighbor of particle $i$. 
In SANN, this symmetry is not ensured, because every particle has its own neighbor shell radius.
Thus the distance between both particles can be smaller than the shell radius of particle $i$ and
larger than that of particle $j$ at the same time; hence, asymmetries can occur. However, we have
found that the fraction of asymmetric neighbors is quite small: below 5\% for the systems
we studied. Moreover, these tend to be those neighbors that are far away, i.e. contribute a small
solid angle and are arguably of minor importance for the neighborhood. Therefore, for many applications
this might not matter. But in case it does we provide two (arbitrary) ways to make the algorithm "symmetric":
if $j$ is a neighbor of $i$ but not vice-versa, either a) remove the asymmetric pair, i.e. remove
$j$ from the list of neighbors of $i$, or b) complete the asymmetric pair, i.e. add $i$ to the list
of neighbors of $j$.


\item[Local volume:]
It is possible to assign a local volume to each particle. The Voronoi algorithm has as an obvious
choice for the local volume the Wigner-Seitz cell, and by construction the sum of all local volumes
adds up to the total system volume. For SANN  one can
think of many different definitions of a local volume, e.g. related to the shell or cutoff radius,
but there is no inherent definition. Consequently, the sum of such local volumes does not
by definition equal the system volume. Again, if it is important to attribute a volume to each particle, we can simply (and somewhat arbitrarily) 
rescale all volumes, such that their sum equals the total volume. 

\item[Independence of space dimension:]
Although designed for three-dimensional space, we point out that the algorithm is valid without
modification for any space-dimension with $d \ge 2$ (and for $d=1$, it is obviously not needed). 
In particular in higher-dimensional space,
e.g. when studying the packing of hyper-spherical particles, easy-to-implement algorithms that
go beyond the fixed-distance cutoff are scarce and SANN might be an attractive procedure.

\item[Next-nearest neighbors:]
In principle, the algorithm can easily be extended to yield a set of next-nearest neighbors, e.g.
neighbor particles with a distance corresponding approximately to the second peak of the
pair correlation function $g(r)$. For this task the algorithm is performed twice as follows:
in the first run, the nearest neighbors are computed without any modifications. Then all these
nearest neighbors are discarded from the list of possible neighbors, and the algorithm is run a
second time. Because the algorithm is scale-free, no modification to the algorithm is required,
and the next-nearest neighbor shell is obtained. Note that simply increasing the total solid
angle to $8 \pi$ in Eqn.~\ref{eqn:sann} does not work, as the solid angle contribution of the
nearest neighbors would dominate due to the large shell radius $R^{(m)}_i$. 
As we shall see later in the paper, in practice this extension does not work particularly well for finding
next-nearest neighbors.


\end{description}

\section{Simulation Details}

Below, we briefly describe the systems and the simulation methods used to produce the test
configurations studied in this paper. Moreover, we provides details about  the library used for the Voronoi
construction and the implementation of our SANN algorithm. At the end of this section we
briefly review the bond-order correlators that we use later to perform a structural
analysis on some of the systems.

In what follows, we denote the temperature by  $T$, the pressure by $P$ and the (number) density by  $\rho$.
The packing fraction is defined as $\phi=\frac{\pi}{6}\rho d^3$, where $d$ is the
particle's diameter. In what follows, $\sigma$ will be the unit of length for both the
hard-sphere and the Lennard-Jones systems whereas for carbon we will express the length
in \AA. All distances presented in the manuscript will be expressed in the appropriate length units.

\subsection{Sample preparation}\label{sec:samples}

Monodisperse hard-sphere configurations were prepared using an event-driven molecular
dynamics simulation in an NVT ensemble with $N=86400$ particles in a cubic box with
periodic boundary conditions and with temperature $T = 1$, mass $m=1$ and diameter $d=1$. The
system was prepared at a packing fraction ($\phi=0.54$) within the solid-liquid coexistence region
($\phi_f=0.492$ and $\phi_s=0.543$~\cite{Noya_2008}) and at a higher packing fraction ($\phi=0.61$)
beyond the hard-sphere glass-transition packing fraction ($\phi_g=0.58$). 
In both configurations 
only 1\% of the particles are labeled as solid-like with a $q_6$ bond order 
criterion; note that the $q_6$ criterion will be discussed later in the text.
For more details on
these simulations, we refer the reader to Refs.~\cite{Sanz_2011} and~\cite{Valeriani_2011}.

The carbon phases were simulated using the LCBOPI\textsuperscript{+} potential~\cite{Ghiringhelli_2008}
at the same conditions as the study on diamond nucleation in
Ref.~\cite{Ghiringhelli_2007}, namely $P = 30 GPa$ and $T = 3750 K$ for the $3$-fold coordinated
liquid and graphite, and $P = 85 GPa$ and $T = 5000 K$ for the $4$-fold coordinated liquid and
diamond phases. Both conditions correspond to $25\%$ under-cooling with a nucleation free-energy
barrier equal to or larger than $\Delta G = 25 k_B T$ (with $k_B$ Boltzmann's constant)
preventing spontaneous crystallization of the metastable liquid phase. All systems contained
$N = 1000$ particles, with the exception of the graphite crystal which had $N = 960$ particles.
More details on the simulation methods and the semi-empirical interaction potential are given
in Ref.~\cite{Ghiringhelli_2008}.

In our discussions of all Lennard-Jones systems, we denote with $T^*$ and $P^*$ the temperature
and pressure in reduced units ($T^*=k_BT/\epsilon$ and $P^*=P\sigma^3/\epsilon$), with $\epsilon$
the Lennard-Jones well-depth, and $\rho^*=\rho \sigma^3$ the density in reduced units. To
construct configurations of the Lennard-Jones fcc crystal and liquid phases we performed
Monte Carlo simulations in the isothermal-isobaric ensemble for particles interacting via
a truncated and shifted Lennard-Jones pair potential~\cite{Frenkel_2002,Errington_2003} with
a cutoff distance of $2.5$. For both phases, a system of $N=4000$ particles
was prepared at the reduced temperature $T^*=0.92$ and pressure $P^*=5.68$. Under these
conditions, which correspond to $20\%$ under-cooling with respect to coexistence,
the liquid phase is metastable with respect to the fcc crystal phase. However, a nucleation
free-energy barrier of $\Delta G \approx 20 k_B T$ prevents spontaneous crystallization
on simulation time scales~\cite{Wolde_1996}. The two-phase liquid-crystal system was simulated
using the same Lennard-Jones potential at the same conditions, but with $N=8000$ particles. The
equilibration was biased with a quadratic potential on the number of solid-like particles to
prevent the further growth of the crystal phase. See Ref.~\cite{Frenkel_2002} for details on
biased Monte Carlo simulations and Ref.~\cite{Wolde_1996a} on how to identify solid-like
particles. The two-phase liquid-vapor configurations were prepared using the same Lennard-Jones 
potential and equilibrated using $NVT$ Monte Carlo simulations at reduced temperature $T^*=1.0$, 
number density $\rho^*=0.3$, and a system size of $N=5000$ particles. In order to study the
liquid-vapor interface, the simulation box was elongated along the x-axis such that the box
length in the x direction was 2.5 times as long as the y and z directions. As a result of
this simulation box geometry, the resulting liquid-gas interface was perpendicular to the
x-axis.

\subsection{Voronoi and SANN implementation details}\label{sec:implementation}

To compute the Voronoi construction, we used the Open Source Computational Geometry Algorithms
Library (CGAL~\cite{Pion_2008}), version 3.7. However, we were only interested in the set of nearest
neighbors and not in the additional Voronoi information such as the volume, faces, edges, etc. of the
Wigner-Seitz cell. Therefore, it was sufficient and computationally cheaper to perform a Delaunay
triangulation, which is the dual of the Voronoi construction. Either construction can be transformed
into the other and both yield identical nearest-neighbor sets. We used CGAL's ``exact predicates inexact
construction'' kernel and included $8$ periodic copies of each particle to emulate periodic boundary
conditions. Although CGAL does support $3d$ Delaunay triangulation with $3d$ periodicity, it turned
out that the run-time was significantly worse. The particles were inserted sequentially to map
CGAL vertex handles and our particle ids.

To speed up the SANN algorithm we made use of a
Verlet list~\cite{Frenkel_2002} with a long cutoff distance to determine the set of possible
neighbors for each particle. Although this method involves a cutoff parameter, we could have
chosen a parameter-free algorithm like a binary space partitioning tree or an octree~\cite{Eberhardt_2010}.
In general, any domain-decomposition method suffices as long as it provides enough particles
for the algorithm to converge.

\subsection{Bond-order correlator}

In order to identify solid-like particles in some of the systems, we used local bond-order
parameters according to Ref.~\cite{Wolde_1996a}. The original order parameter
described by Steinhardt {\it et al.}~\cite{Steinhardt_1983} is based on the idea
of expanding the neighborhood of each particle in a system in terms of a specific
set of spherical harmonics, e.g. expanding in terms of the spherical harmonics with
$l=3$, $l=4$ or $l=6$, depending on the local symmetry. The algorithm was later refined
by ten Wolde {\it et al.}~\cite{Wolde_1996a} for the study of nucleation, and has proven
to be a useful tool even in the case of higher-dimensional systems~\cite{Meel_2009a,Meel_2009d}.

To compute the bond-order parameter each particle $i$ is assigned a $(2l + 1)$-dimensional
complex vector $\vec{q}_l(i)$ whose $m$-th component is defined by,
\begin{equation}\label{eqn:bond}
q_l^m(i) = \frac{1}{N_b(i)} \sum_{j} Y_{lm}(\vec{\hat{r}}_{ij} ),
\end{equation}
where $N_b(i)$ denotes the number of nearest neighbors, $Y_{lm}(\vec{\hat{r}}_{ij})$ is the
set of spherical harmonics of order $l$ with components $-l \le m \le l$, $\vec{\hat{r}}_{ij}$ is
the unit vector pointing from the center of $i$ to its neighbor $j$, and the sum runs over
all neighbors $\{ j \}$ of particle $i$. From this we can construct a measure for the
neighborhood similarity of two particles,
\begin{equation}
d_l(i,j) = \frac{\vec{q}_l(i) \cdot \vec{q}_{l}^*(j)}{|\vec{q}_l(i)| \, |\vec{q}_l(j)|},
\end{equation}
where the superscript star denotes the complex conjugate. We call the $d_l(i,j)$ the local
bond-order correlator, which is one when both particles are in an identically ordered
environment. To distinguish reliably between solid-like and liquid-like particles, particularly
in an under-cooled liquid, additional steps are required to increase the contrast. However,
since a change in the neighborhood algorithm already affects this stage of the analysis, we
will not follow the procedure to the end, but instead compare the local bond-order correlators. 

\section{Results}\label{sec:results}

In what follows we apply the proposed algorithm (SANN) to several simulation samples
and compare the resulting set of nearest neighbors to the sets obtained from both the
fixed-distance cutoff criterion and the Voronoi construction. Moreover, on some systems
we perform a structural analysis using bond-order parameters and discuss the impact 
different nearest-neighbor sets have on the bond-order correlator distributions. We
finish by presenting run-times of each algorithm for several simulation samples.

\subsection*{Bulk phases}

To start, we compute the nearest-neighbor distribution $P(N_n)$ for the bulk phases
described in Section~\ref{sec:samples} using three neighborhood algorithms.
For the fixed-distance cutoff, we set the cutoff to the minimum of the pair
correlation function, which yields $r_c = 1.5$ and $r_c = 1.35$ for the
Lennard-Jones liquid and fcc crystal phases, respectively, $r_c = 2.0$ for
all carbon phases, and $r_c = 1.35$ and $r_c = 1.3$ for the low- and high-density
hard-sphere suspensions.

\begin{figure*}
\center
a)\includegraphics[width=0.45\textwidth]{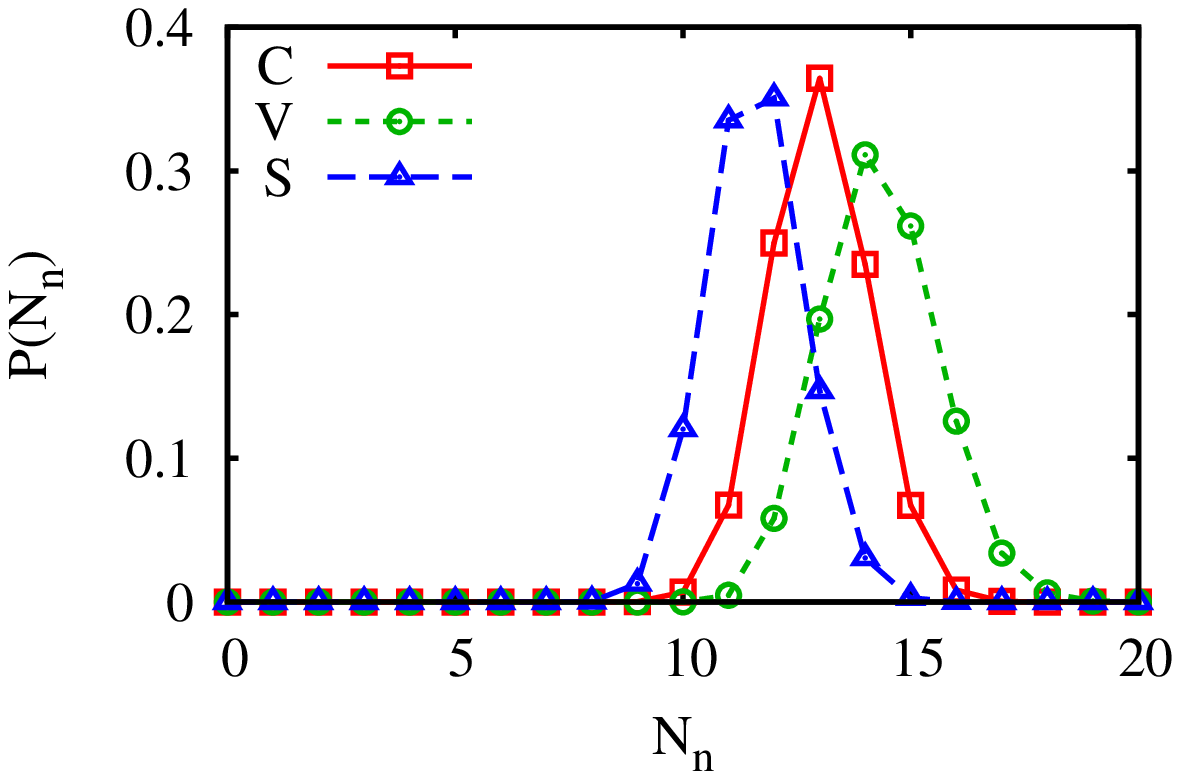}
b)\includegraphics[width=0.45\textwidth]{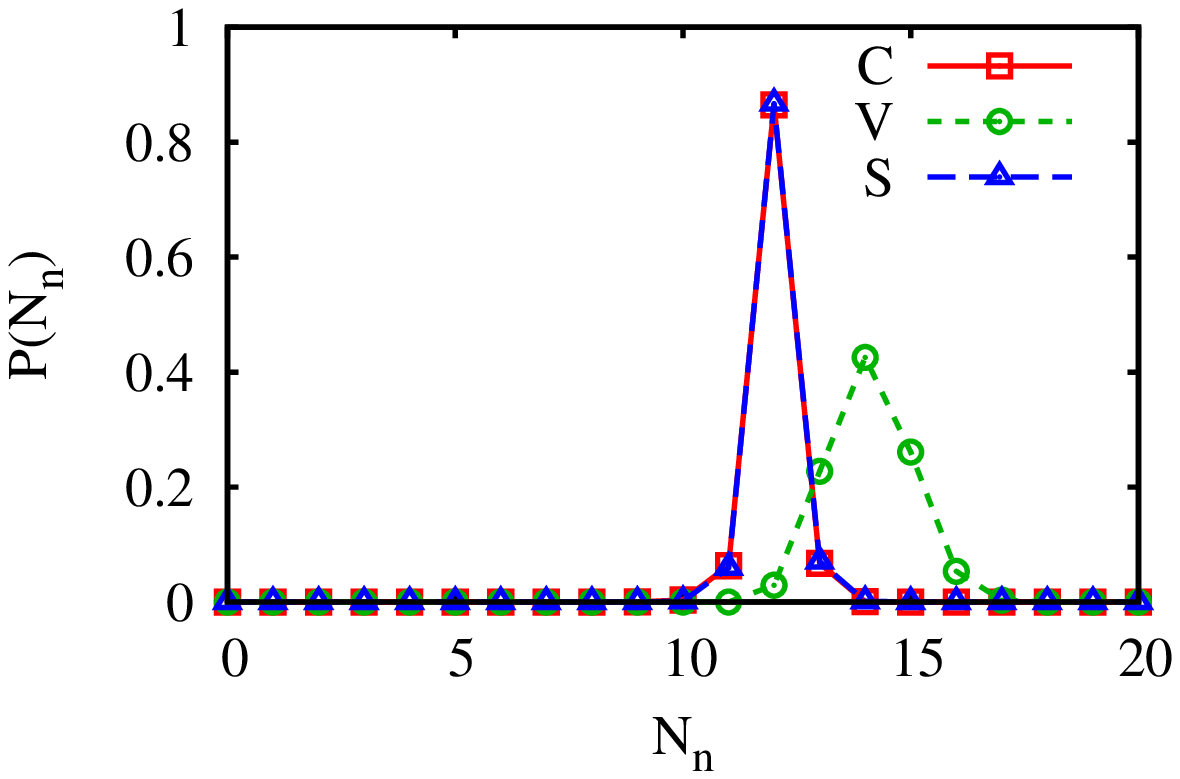}
c)\includegraphics[width=0.45\textwidth]{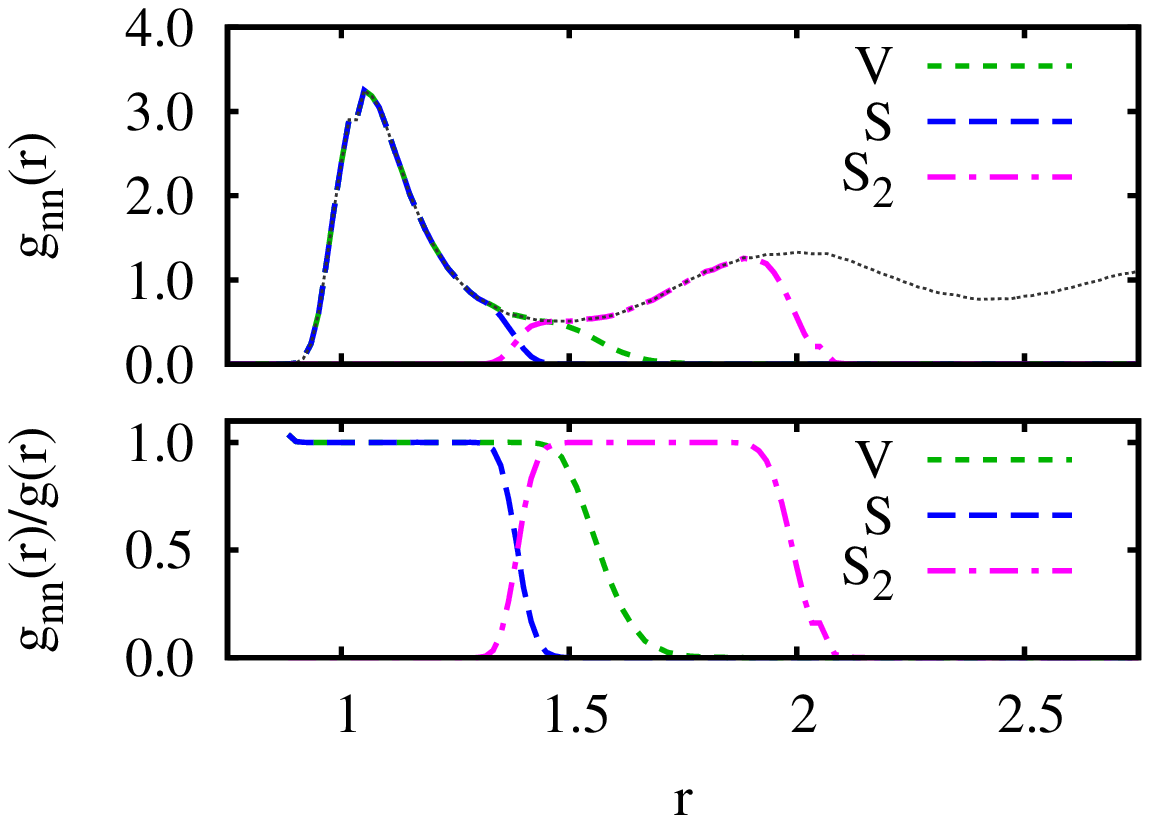}
d)\includegraphics[width=0.45\textwidth]{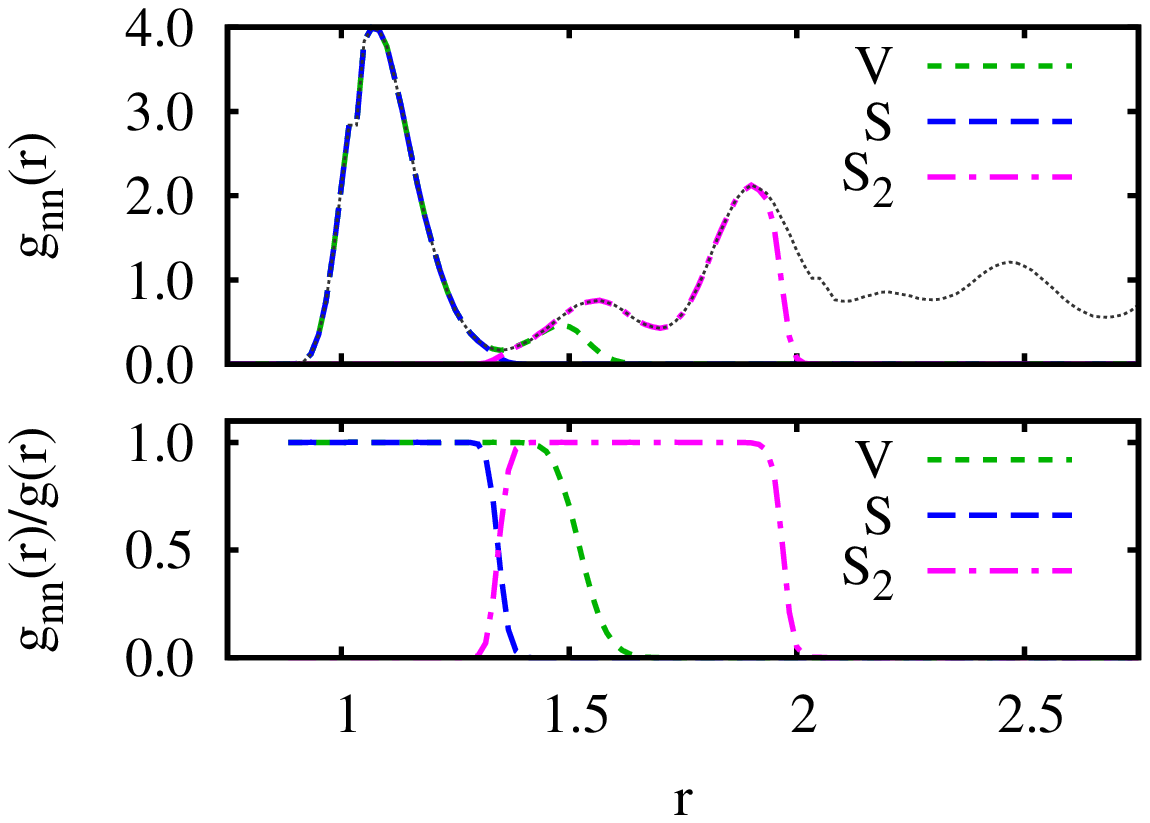}
\caption{Nearest neighbors distribution $P(N_n)$ for a Lennard-Jones liquid (panel a) and
fcc crystal (panel b) obtained by fixed-distance cutoff ($C$), Voronoi construction ($V$) and
SANN considering neighbors belonging to the first coordination shell ($S$). Panels c) and d)
plot the pair correlation functions $g(r)$, considering all particles, as a reference (thin
grey dotted line), and $g_{nn}(r)$, considering only nearest neighbors (Voronoi ($V$)
and SANN ($S$)) and next-nearest neighbors (SANN ($S_2$)), for both the liquid (c) and the
fcc crystal (d). In addition, their fraction $g_{nn}(r) / g(r)$ is shown.}
\label{fig:lj}
\end{figure*}

Figures~\ref{fig:lj}a and~\ref{fig:lj}b depict for the liquid and fcc Lennard-Jones
system the nearest-neighbor distribution $P(N_n)$ computed using fixed-distance cutoff distance
($C$), Voronoi construction ($V$) and our SANN algorithm for nearest neighbors ($S$) and for 
next-nearest neighbors ($S_2$). In
both systems, the Voronoi construction identifies more nearest neighbors on average
than the fixed-distance cutoff and SANN methods. Its peak in the nearest-neighbor
distribution is around $14$ neighbors for both the metastable-liquid and the fcc phases.
The fixed-distance cutoff ($C$) exhibits a nearest-neighbor distribution which peaks
around $13$ neighbors in the liquid and at $12$ in the fcc crystal, whereas the distribution
obtained using the SANN algorithm peaks around $11-12$ neighbors for the liquid and
sharply at $12$ neighbors in the fcc crystal. Note that $12$ is also the number that
one would expect from a close-packed arrangement of spheres.

In order to get a better understanding regarding the particles which are identified
as nearest neighbors in the SANN and Voronoi algorithms, we also compute the pair
correlation function using only the nearest neighbors obtained with each method ($g_{nn}(r)$)
and compare them to the $g(r)$ of all particles. In the following discussions, we
picture the environment around each particle to consist of several shells, each shell
associated with a peak in the $g(r)$. Hence, everything up to the first minimum of the 
$g(r)$ corresponds to the first shell, everything between the first and the second minimum to
the second shell and so forth. The fixed-distance cutoff method is not applied here as, 
by definition, its $g_{nn}(r)$ yields $g(r)$ exactly up to the cutoff radius, after which
it is zero. The upper graphs 
show the pair correlation functions $g_{nn}(r)$ and $g(r)$
for reference, and the lower graphs show the ratio $g_{nn}(r) / g(r)$. At a given distance
$r$, the latter ratio gives 1 if all particles at this distance are identified as nearest
neighbors, and reduces to zero if none of these particles are considered neighbors.
Hence, a steep decrease in the ratio $g_{nn}(r) / g(r)$ indicates few fluctuations 
in the selection of the neighbors. In addition to the nearest neighbors, the graphs 
also show results for the next-nearest neighbors obtained from the SANN method ($S_2$).

Figure~\ref{fig:lj}c and~\ref{fig:lj}d plots these functions for the Lennard-Jones phases.
They show that $g_{nn}(r)$ for the Voronoi construction ($V$) is identical to the reference
$g(r)$ up to the first minimum, and in the fcc crystal even slightly beyond that. From the
position of the decrease in the $g_{nn}(r) / g(r)$, i.e. slightly to the right of the first
minimum in the $g(r)$, we see that the Voronoi algorithm also includes some particles from
the second neighbor shell (see upper panels of Figs.~\ref{fig:lj}c and ~\ref{fig:lj}d).
This behavior originates from fluctuations which cause the Voronoi cell of next-nearest
neighbors to occasionally share a small face~\cite{Troadec_1998}. There exist extensions
to the Voronoi construction which attempt to increase the robustness of the algorithm to
fluctuations~\cite{Brostow_1998,Bandyopadhyay_2004,Yu_2005}. However, many of them introduce
non-inherent parameters and, as such, are not parameter-free. Therefore, we will not consider
them here. In contrast to the Voronoi algorithm, the $g_{nn}(r)/g(r)$ associated with the SANN
algorithm ($S$) drops to zero at the first minimum for both the liquid and crystal phases and
therefore hardly includes any next-nearest neighbors. The SANN algorithm to determine next-nearest
neighbors, denoted $S_2$, does not yield very precise results. 
In particular, in the liquid $S_2$ finds a few spurious particles from the first neighbor
shell and only a fraction of next-nearest neighbors, and in the solid, where it does identify
all next-nearest neighbors, it also includes a considerable amount from the third neighbor shell.
In both cases this can be attributed to the form of the $g(r)$, i.e. the broadness of the second
peak in the liquid, and the closeness of the second and third peaks in the solid. Unfortunately,
this is a recurrent problem with trying to use SANN to determine next nearest neighbors, and for
this reason $S_2$ will not be discussed further in this paper.

\begin{figure*}
a)\includegraphics[width=0.45\textwidth]{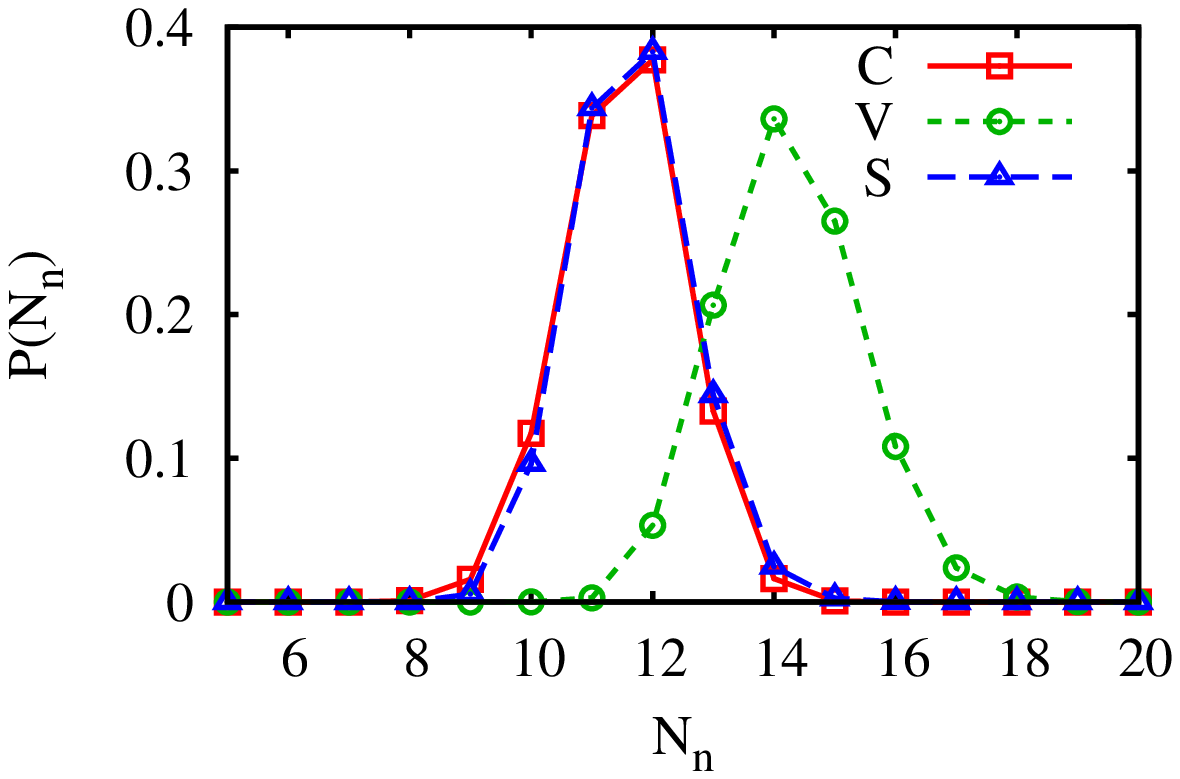}
b)\includegraphics[width=0.45\textwidth]{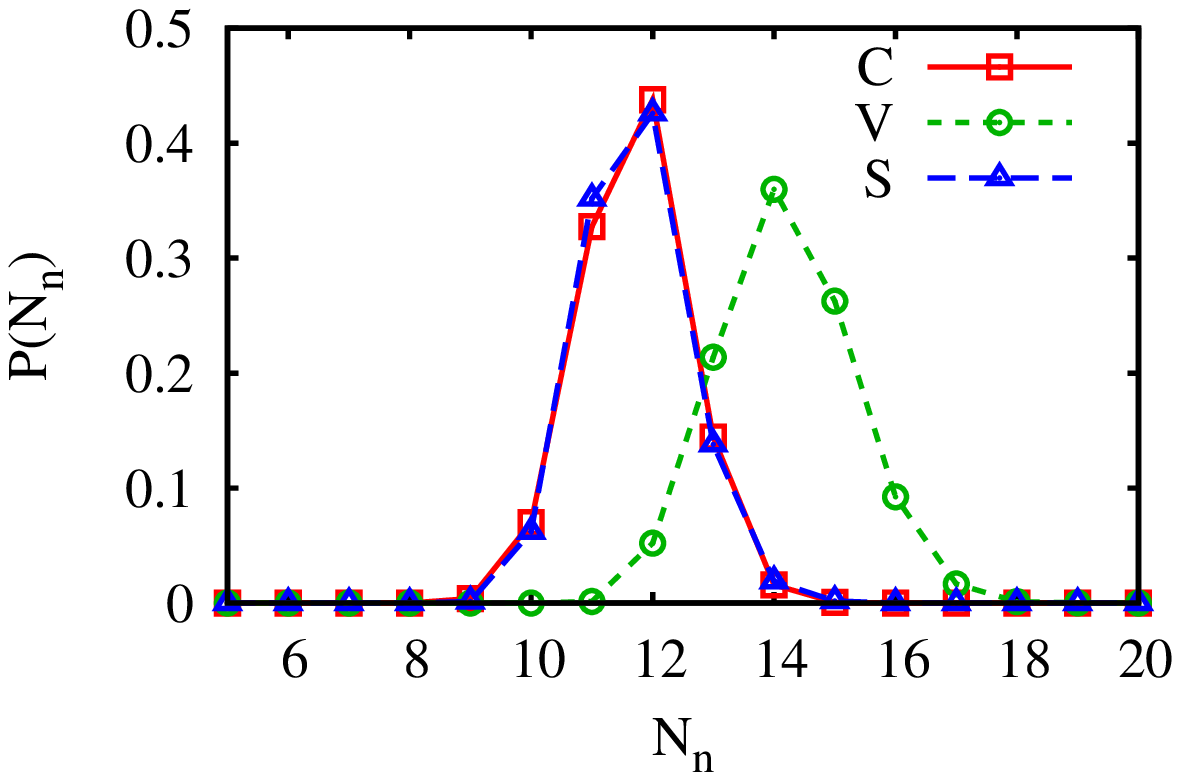}
c)\includegraphics[width=0.45\textwidth]{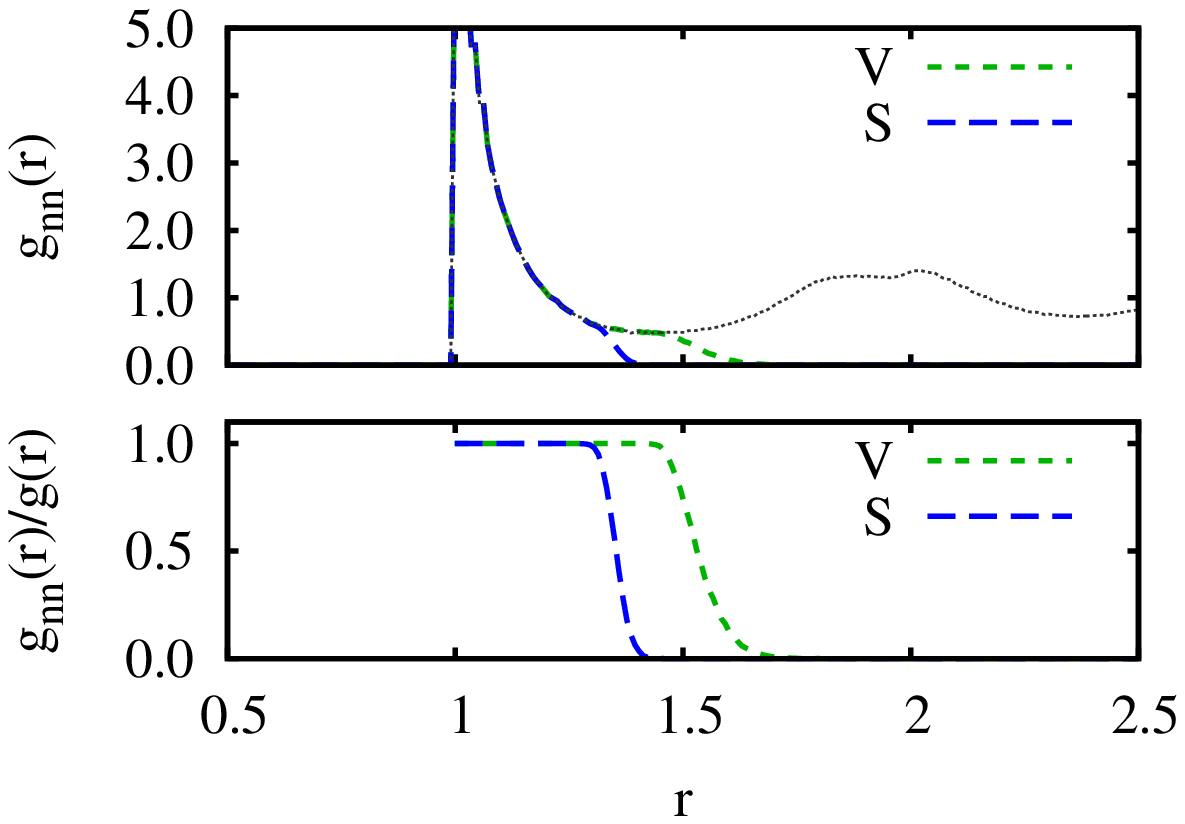}
d)\includegraphics[width=0.45\textwidth]{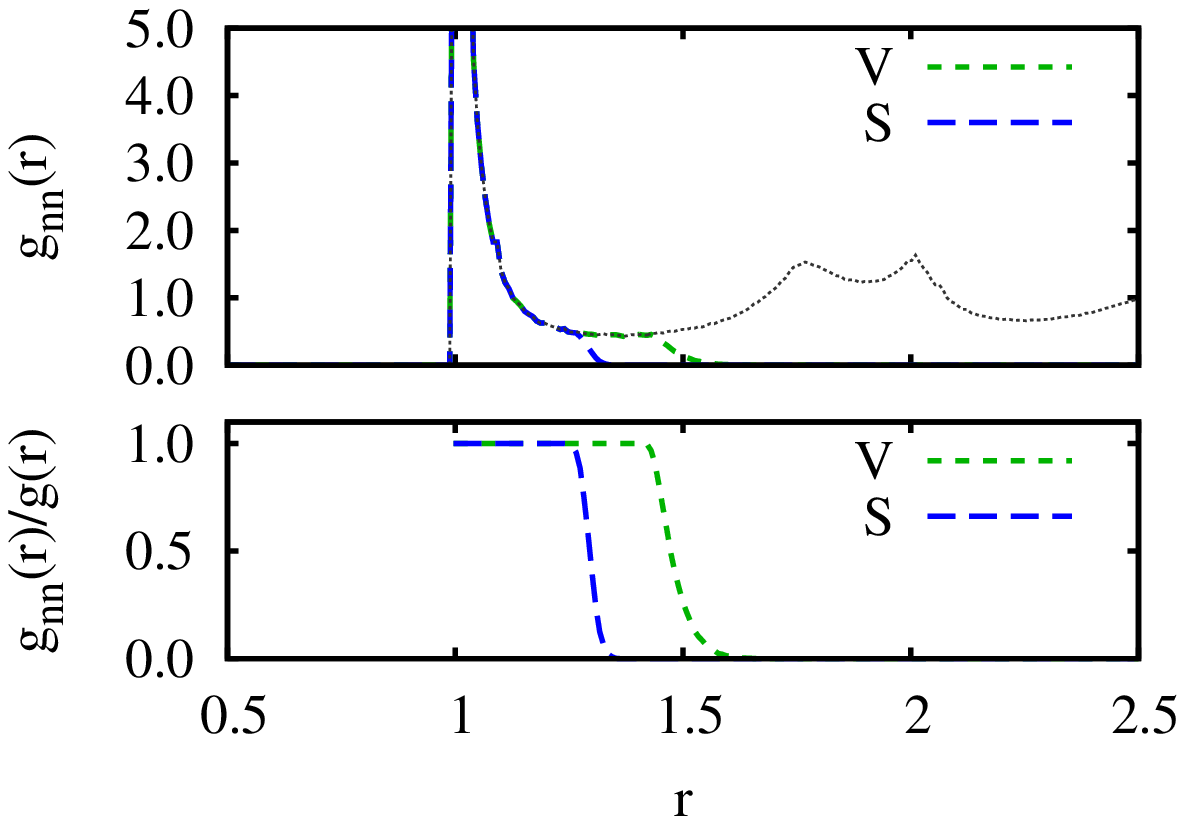}
\caption{Nearest-neighbor distribution and $g(r)$ 
as in Figure~\ref{fig:lj}, but for a monodisperse hard spheres system at
$\phi=0.54$ (panels a and c) and $\phi=0.61$ (panels b and d).
}
\label{fig:hs}
\end{figure*}

For both hard-sphere systems studied, i.e. $\phi = 0.54$ (fluid) and $\phi = 0.61$ (glass), 
the nearest-neighbor distributions of Figures~\ref{fig:hs}a and~\ref{fig:hs}b computed
using the Voronoi construction ($V$) present a peak around $14$ neighbors, given that
some of the neighbors from the second shell are included (as shown in Figs.~\ref{fig:hs}c
and~\ref{fig:hs}d). In contrast, the distribution obtained using the fixed-distance cutoff
($C$) algorithm and the one obtained with SANN ($S$) are fairly similar (Figs.~\ref{fig:hs}a
and~\ref{fig:hs}b) and both peaked around $12$ neighbors for both packing fractions.
Again, this is the number one would expect in a close-packed arrangement of spherical particles.
From Figures~\ref{fig:hs}c and~\ref{fig:hs}d we see that the pair correlation function
for the Voronoi construction is identical to the reference $g(r)$ up to the first minimum.
But, as in the Lennard Jones system, it also seems to partially include particles from the
second neighbor shell (see upper panels of Figs.~\ref{fig:hs}c and d). In contrast, the
$g_{nn}(r)$ computed using SANN drops to zero at the first minimum at both $\phi$'s
and does not include next-nearest neighbors.

\begin{figure*}
a)\includegraphics[width=0.45\textwidth]{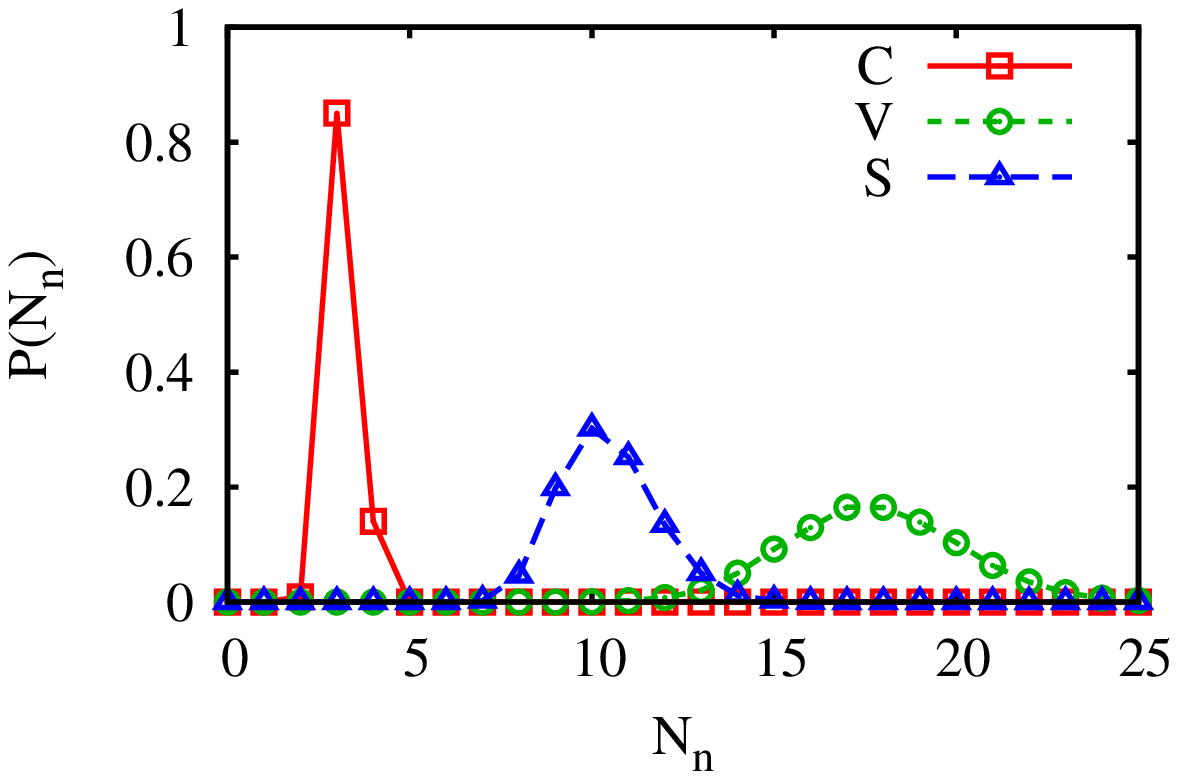}
b)\includegraphics[width=0.45\textwidth]{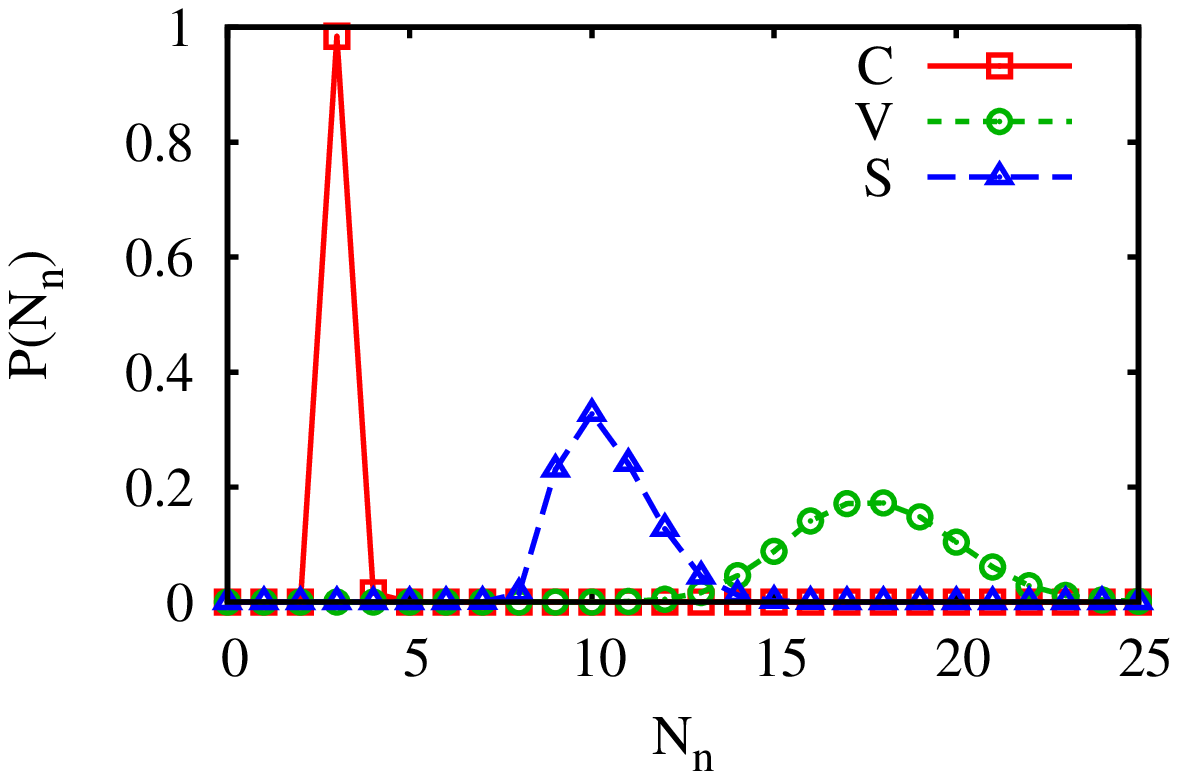}
c)\includegraphics[width=0.45\textwidth]{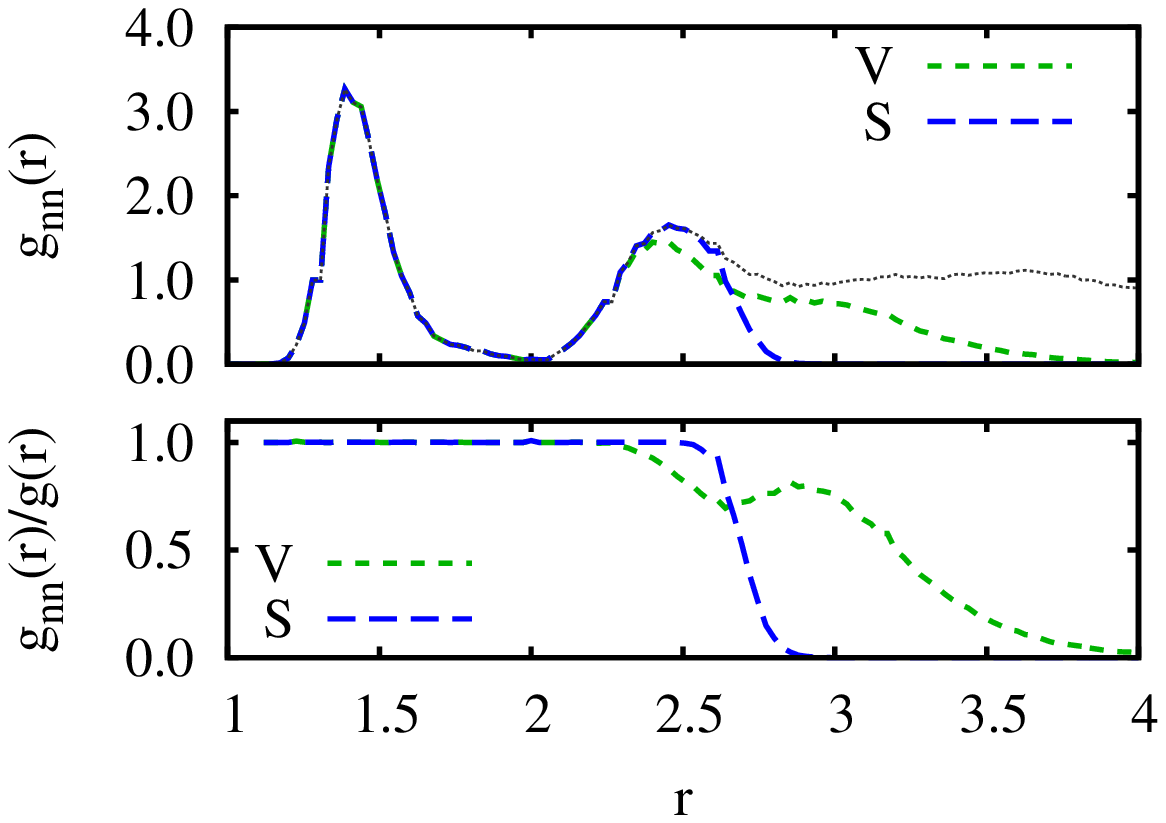}
d)\includegraphics[width=0.45\textwidth]{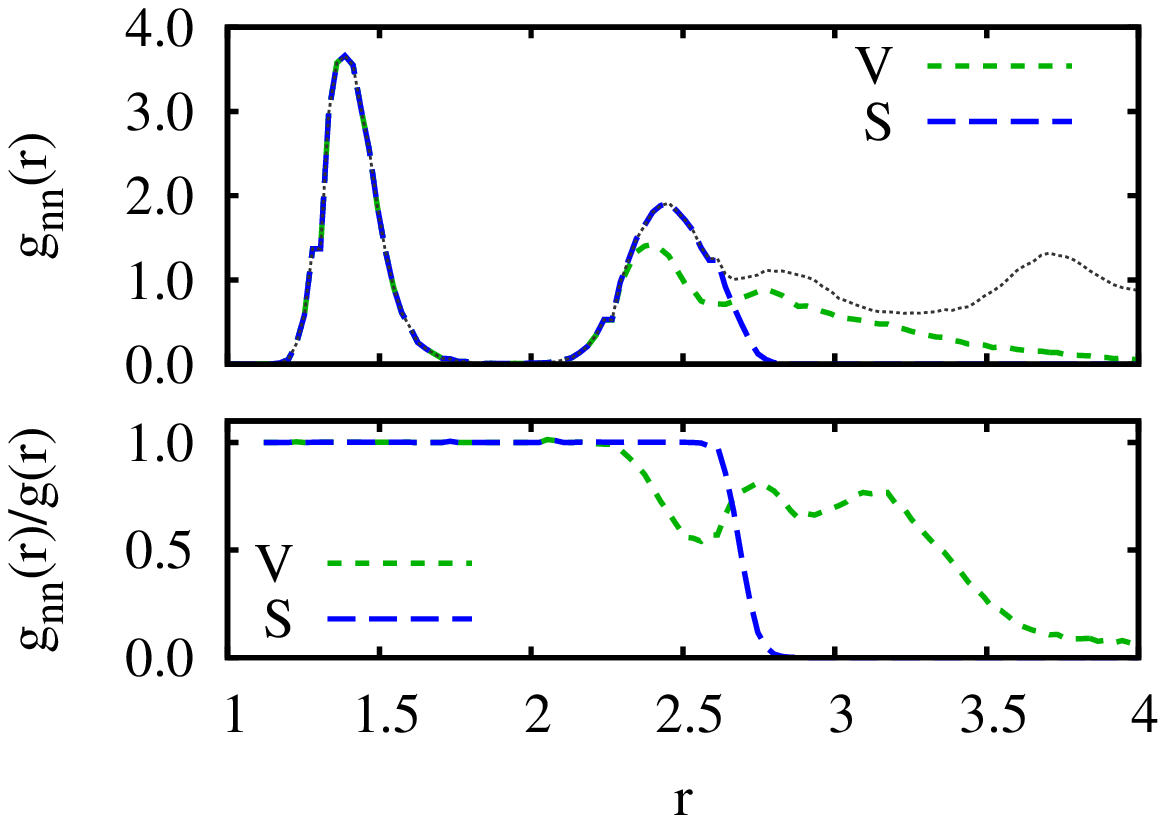}
\caption{Nearest-neighbor distribution and $g(r)$
as in Figure~\ref{fig:lj}, but for both a
$3$-fold coordinated carbon liquid (panels a and c) and graphite crystal (panels b and d).
}
\label{fig:gra}
\end{figure*}

\begin{figure*}
a)\includegraphics[width=0.45\textwidth]{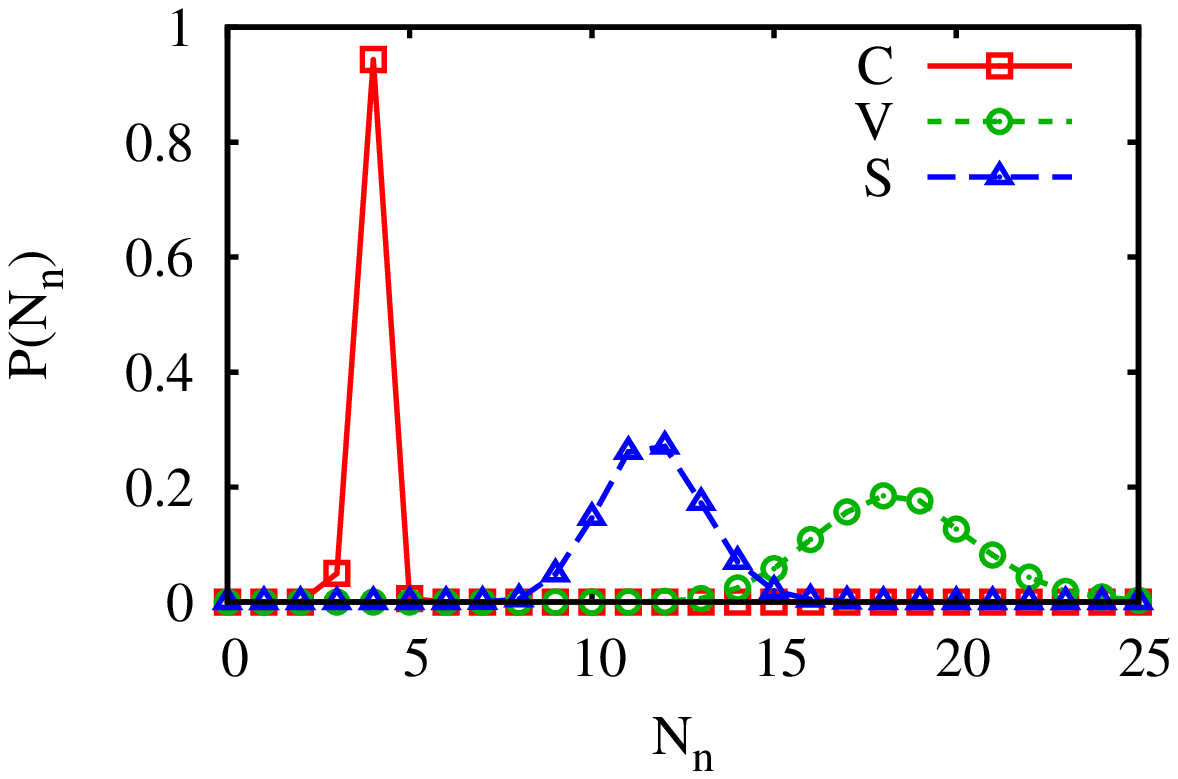}
b)\includegraphics[width=0.45\textwidth]{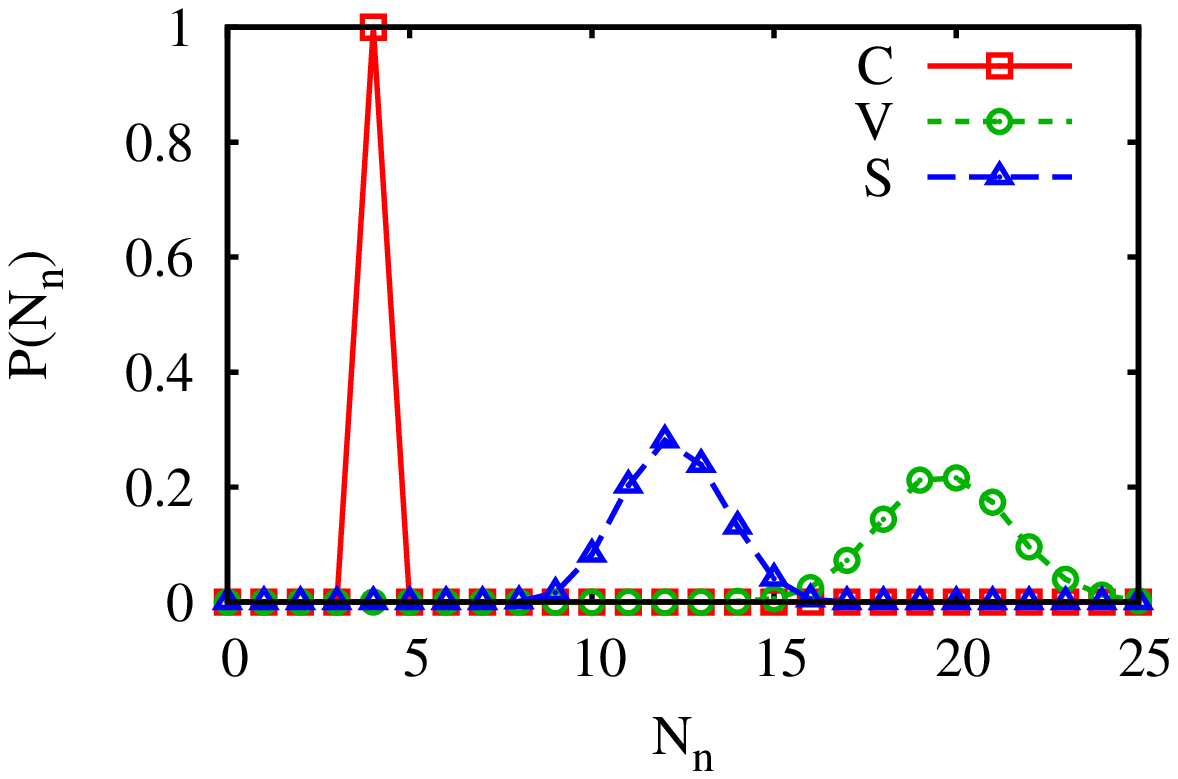}
c)\includegraphics[width=0.45\textwidth]{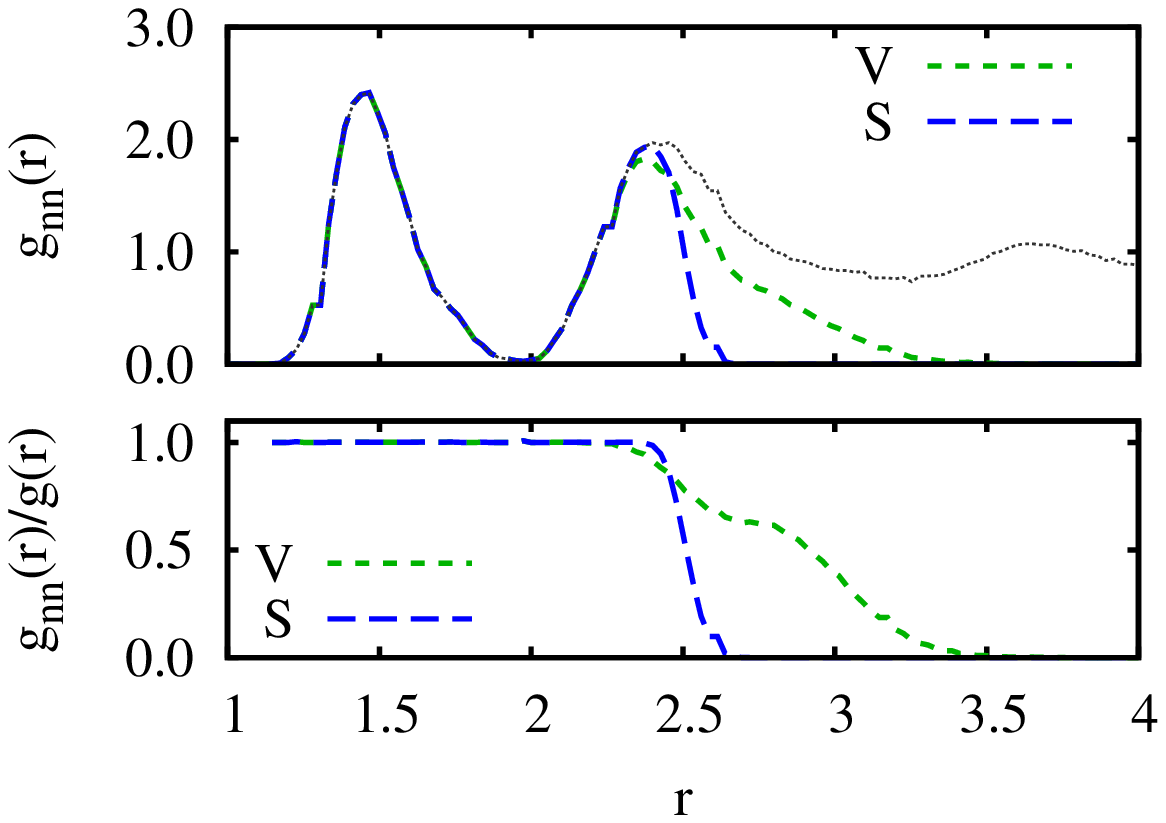}
d)\includegraphics[width=0.45\textwidth]{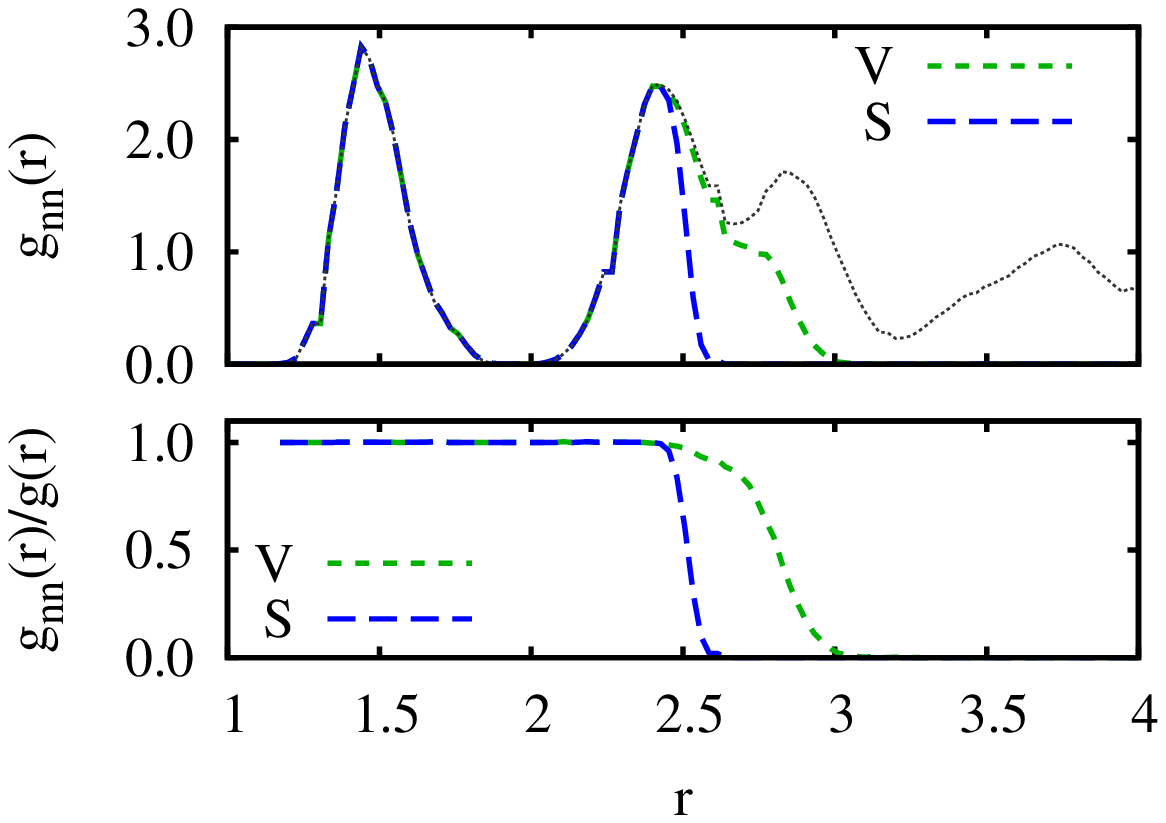}
\caption{Nearest-neighbor distribution and $g(r)$ as in Figure~\ref{fig:lj}, but for both a
$4$-fold coordinated carbon liquid (panels a and c) and diamond crystal (panels b and d).
}
\label{fig:dia}
\end{figure*}

Figures~\ref{fig:gra} and~\ref{fig:dia} depict results for systems consisting of 3-fold
coordinated liquid carbon and graphite and 4-fold coordinated liquid carbon and diamond,
respectively. In contrast to the Lennard-Jones and hard-sphere systems, carbon 
is a highly structured network-forming system, even in the liquid phase. 
It features open structures with few (3 or 4) close-by ordered neighbors: 
the 3-fold coordinated liquid carbon has a graphite-like structure in the first coordination shell,  
whereas the 4-fold coordinated liquid carbon has  
a rather pronounced diamond-like structure in the first coordination shell, 
shown in the strongly anisotropic angular distribution of the first neighbors.
This is reflected in the pair correlation functions $g(r)$ of Figures~\ref{fig:gra}c and~\ref{fig:dia}c, 
that show a sharp first peak followed by a broad deep minimum: a sign that  
up to the second neighbor shell the liquid has a structure almost as pronounced as the one of the 
corresponding solid.	

The upper panels of Figures~\ref{fig:gra} and~\ref{fig:dia} represent the $P(N_n)$ 
computed for the 3-fold coordinated and 4-fold coordinated carbon phases, respectively.
The nearest-neighbor distributions from the three methods differ significantly: with the cutoff
distance set to the first minimum of the $g(r)$, the fixed-distance cutoff yields a distribution
peaked sharply around $3$ (Fig.~\ref{fig:gra},  3-fold coordinated system) and $4$ (Fig.~\ref{fig:dia},
4-fold coordinated system) particles both for the liquid and the crystal phases. Counting particles
that form chemical bonds, the 3-fold coordinated carbon should have $9$ neighbors within the first
two shells ($3$ in the first shell, separated by a distance of about 1.4 \AA, and $6$ in the second
shell belonging to the same graphite layer), whereas the 4-fold coordinated carbon should have $16$
neighbors within the first two shells ($4$ in the first shell, separated by a distance of about 1.54 \AA,  
and $12$ in the second shell). The SANN algorithm peaks around $10$ (3-fold coordinated systems,
Fig.~\ref{fig:gra}) and $12$ (4-fold coordinated systems, Fig.~\ref{fig:dia}) particles for both
liquid and crystal phases. Those numbers indicate that SANN includes in each particle's neighbor list
most of the particles belonging to the second coordination shell, as confirmed by the $g(r)$ plots in the
lower panels of Figures~\ref{fig:gra} and ~\ref{fig:dia}. The Voronoi construction peaks around $17$
(3-fold coordinated systems, Fig.~\ref{fig:gra}) and $20$ (4-fold coordinated systems,
Fig.~\ref{fig:dia}) since it includes particles from the second and third shells. This happens in the
liquids, graphite and diamond, as shown by the non-monotonic decay of the Voronoi's  $g_{nn}(r)/g(r)$
that extends well beyond the second minimum in these cases. 

To explain the behavior for both the Voronoi construction and the SANN method, we recall that both
$3$-fold and the $4$-fold coordinated carbon phases are network-forming open structures. Moreover,
graphite forms layers that are several particle diameters apart (with $3.4$ \AA~ the distance between
two layers). This structure affects both algorithms differently; by definition the Voronoi construction
searches for neighbors that surround the center particle in all space dimensions, attempting to construct
a complete ($3d$) Wigner-Setz cell. For the $4$-fold coordinated carbon phases the Wigner-Seitz cell of the
neighbors from the first coordination shell is a fragile tetrahedron, meaning that it is very likely that
much further apart particles share a small face. Hence it contains particles from the second and even higher
coordination shells. The planar arrangement of neighbors in the $3$-fold coordinated carbon phases forces
the Voronoi construction to consider particles from neighboring layers to complete a $3d$ Wigner-Seitz cell;
particles that, as one might argue, belong to an entirely different neighborhood. In contrast, SANN does
not attempt to complete a $3d$ environment. However, in order to complete its neighborhood with only neighbors
from the first coordination shell they must have almost identical distances. This is rarely the case
in physical systems and therefore it includes more distant particles as well. Then, however, its neighborhood
is dominated by the particles from the first coordination shell, since they are much closer and consequently
contribute much larger solid angles in comparison to more distant particles from the second shell. This effect
seems to be more pronounced in the diamond crystal phase than in the other carbon phases, and it is the reason
why not all particles from the second shell are included. 

\begin{figure*}
a)\includegraphics[width=0.25\textwidth]{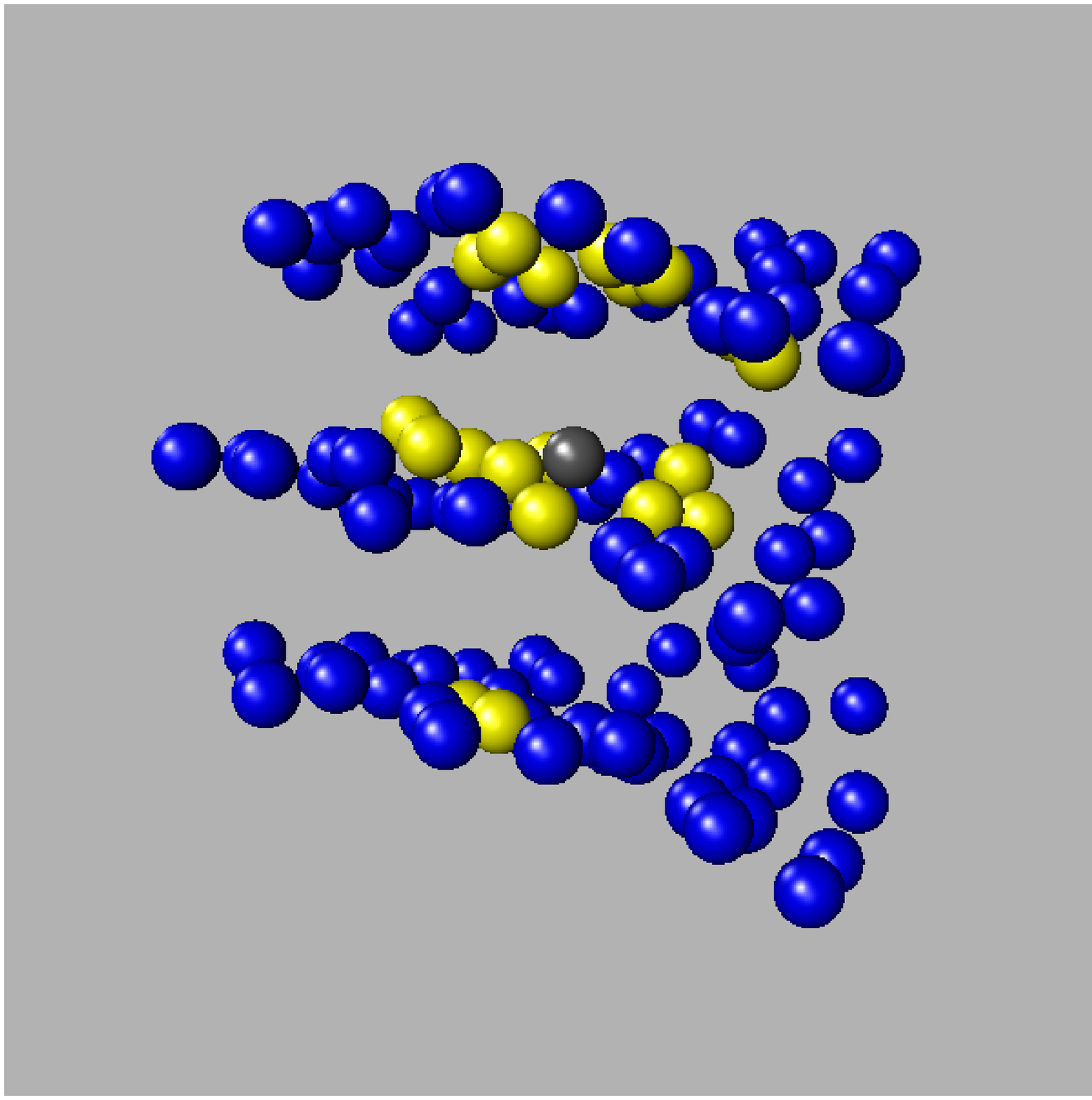}
b)\includegraphics[width=0.25\textwidth]{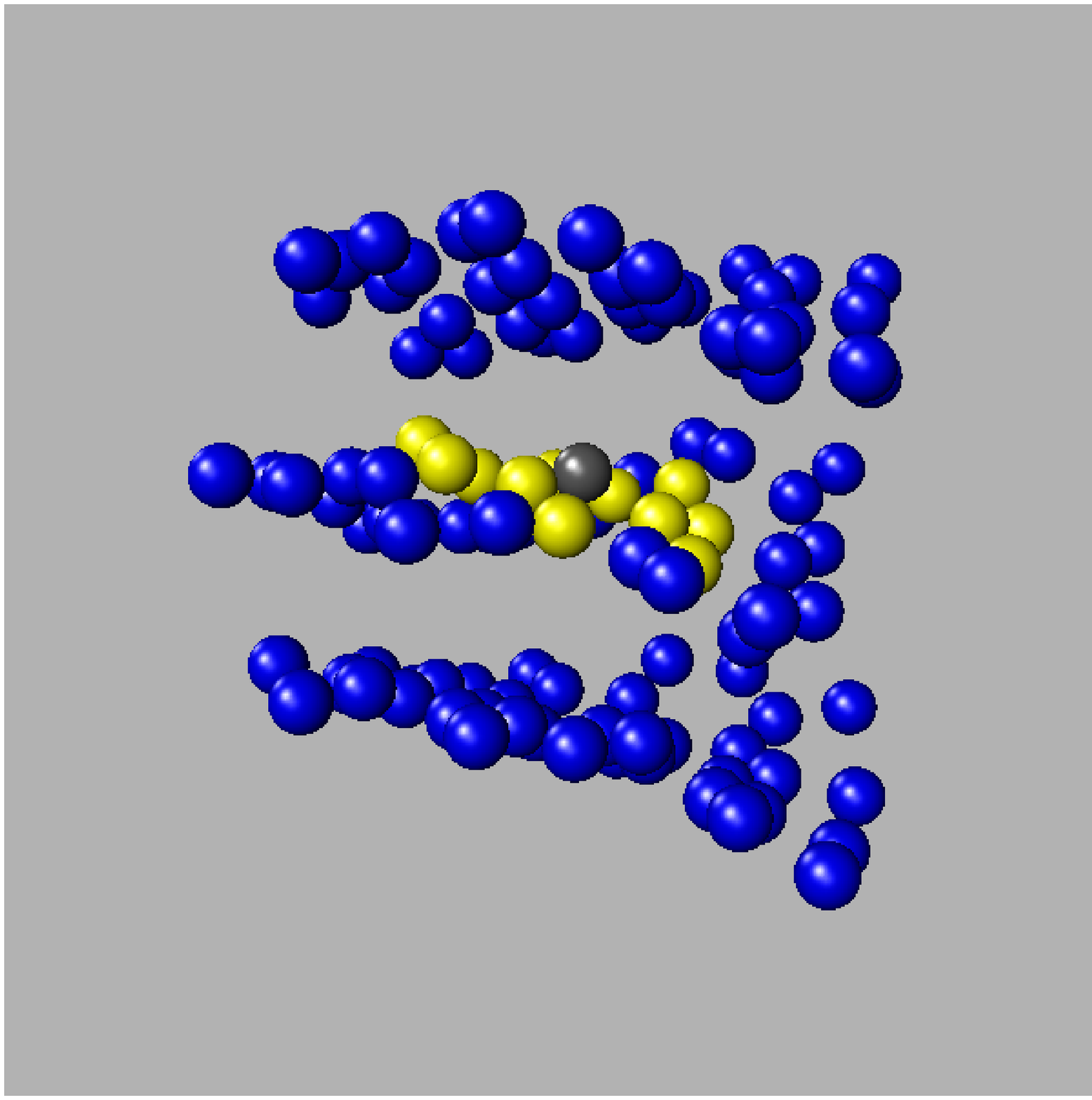}
c)\includegraphics[width=0.25\textwidth]{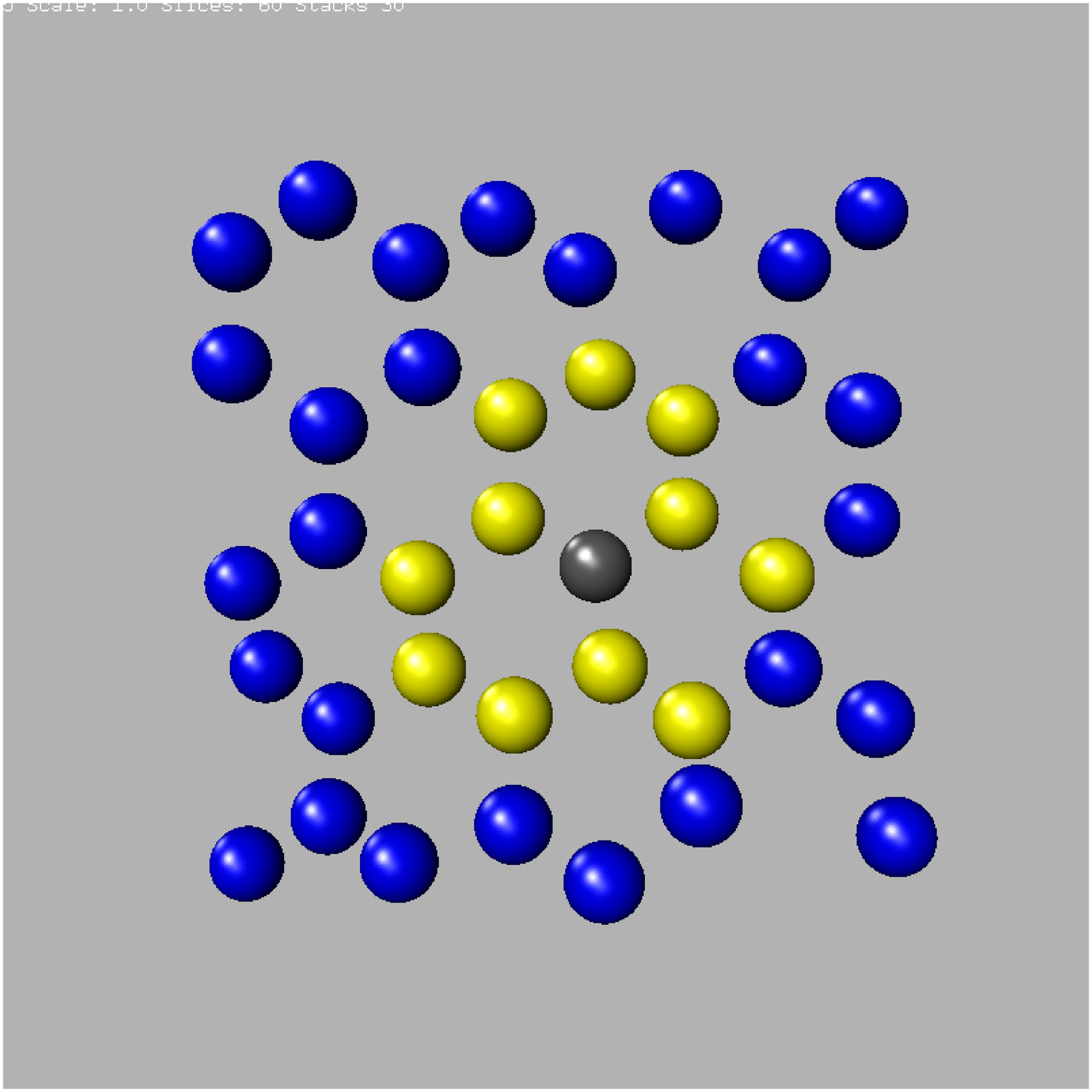}
\caption{Simulation snapshot of 3-fold coordinated carbon graphite showing the first
neighbors (yellow) of a center particle (gray). Surrounding particles that are not 
part of the neighborhood are shown in blue. a) Voronoi construction; 
b) SANN algorithm; c) Top-view on the center layer of panel b.
}
\label{fig:graphite-slab}
\end{figure*}

As an example for this behavior, we depict a few graphite layers in Figure~\ref{fig:graphite-slab}, where
we color particles identified as neighbors with the Voronoi (panel a) and SANN (panel b) algorithms. 
Although geometrically correct, the additional neighbors from arguably different neighborhoods identified
by the Voronoi construction may distort results for local quantities. Figure~\ref{fig:graphite-slab}c presents
a top-view of the center layer of panel (b) and shows that the neighborhood identified by SANN includes the
complete first and almost complete second neighbor shell.

\subsection*{Interfaces}

We now apply the three algorithms to two-phase systems with planar interfaces, namely a
liquid-crystal and a liquid-vapor Lennard-Jones systems as described in
Section~\ref{sec:samples}. The two phases are arranged in a slab-geometry such that
the interfaces are normal to the $x$-direction.  For the fixed-distance cutoff we use $r_c=1.5$,
which corresponds to the minimum of the $g(r)$ for the liquid in the liquid-crystal system;
note that the density of this liquid is not the same as the density of the liquid in the liquid-gas
system. 
This choice of $r_c$ is arbitrary since there is no way to choose a cutoff which satisfies 
all four phases simultaneously.
\begin{figure*}
\center
a)\includegraphics[width=0.45\textwidth]{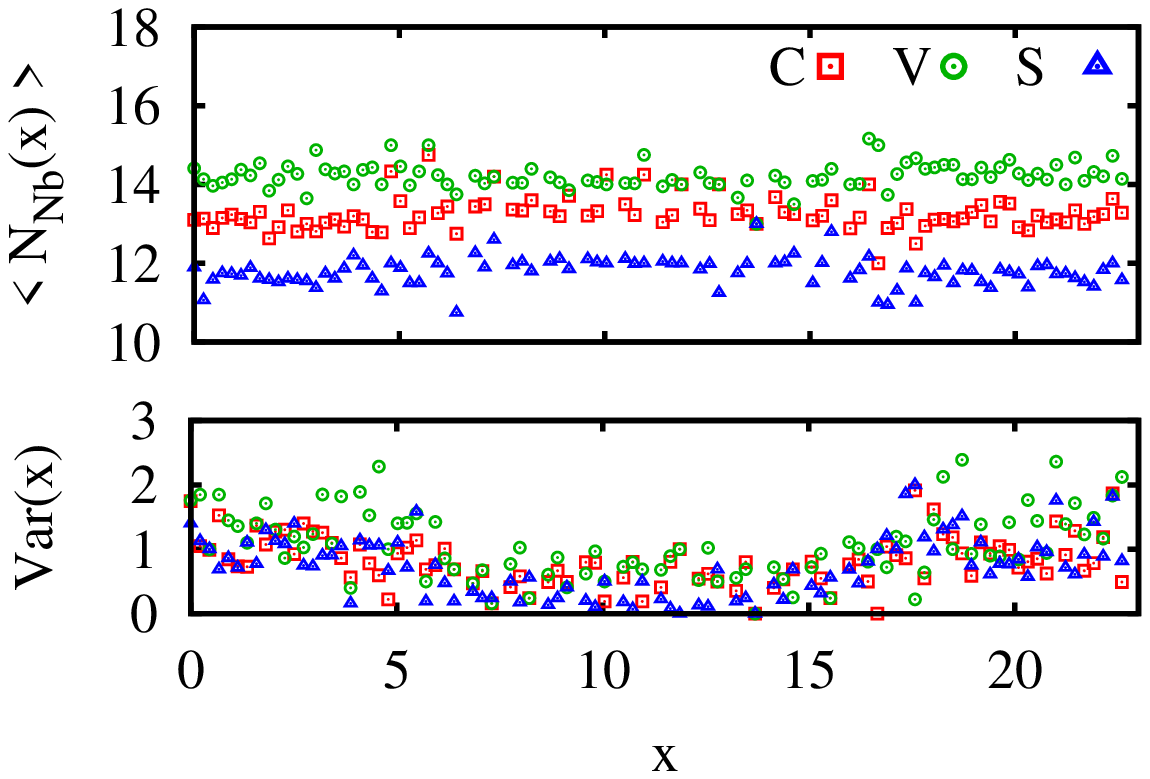}
b)\includegraphics[width=0.45\textwidth]{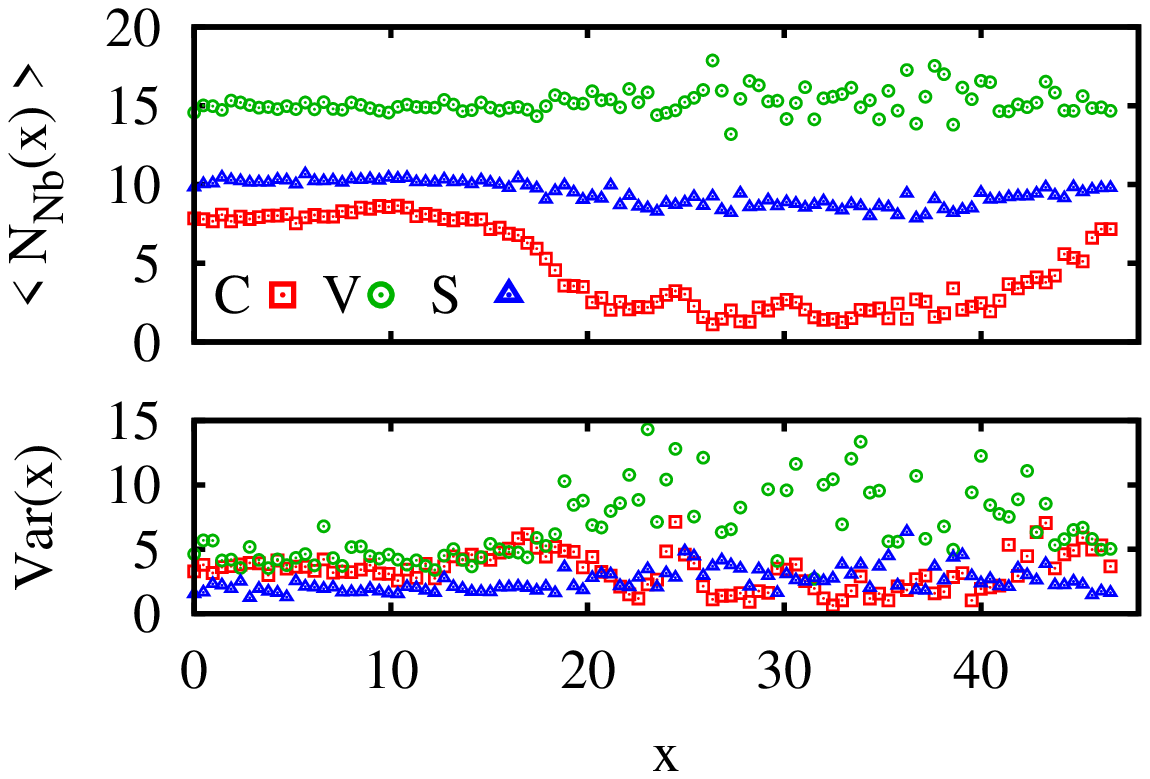}
\caption{Results for two-phase samples in a slab-geometry, with the two interfaces
oriented normal to the $x$-direction. The two phases are a) liquid-crystal and b)
liquid-vapor. Note that the densities of the liquid in both cases are not the same.
For both samples the upper panel shows, as function of $x$-position,
the average number of nearest neighbors, $\Big< N_{Nb}(x) \Big>$, and the lower 
panel the corresponding variance, 
$Var(x) = \Big< N_{Nb}^2(x) \Big> - \Big< N_{Nb}(x) \Big>^2$, for each of the
algorithms: fixed-distance cutoff ($C$) with $r_c=1.5$, Voronoi construction ($V$) and SANN
considering neighbors belonging to the first coordination shell ($S$).  Note the different scales 
on the y-axis.}
\label{fig:tp}
\end{figure*}

\begin{figure}
\includegraphics[width=0.45\textwidth]{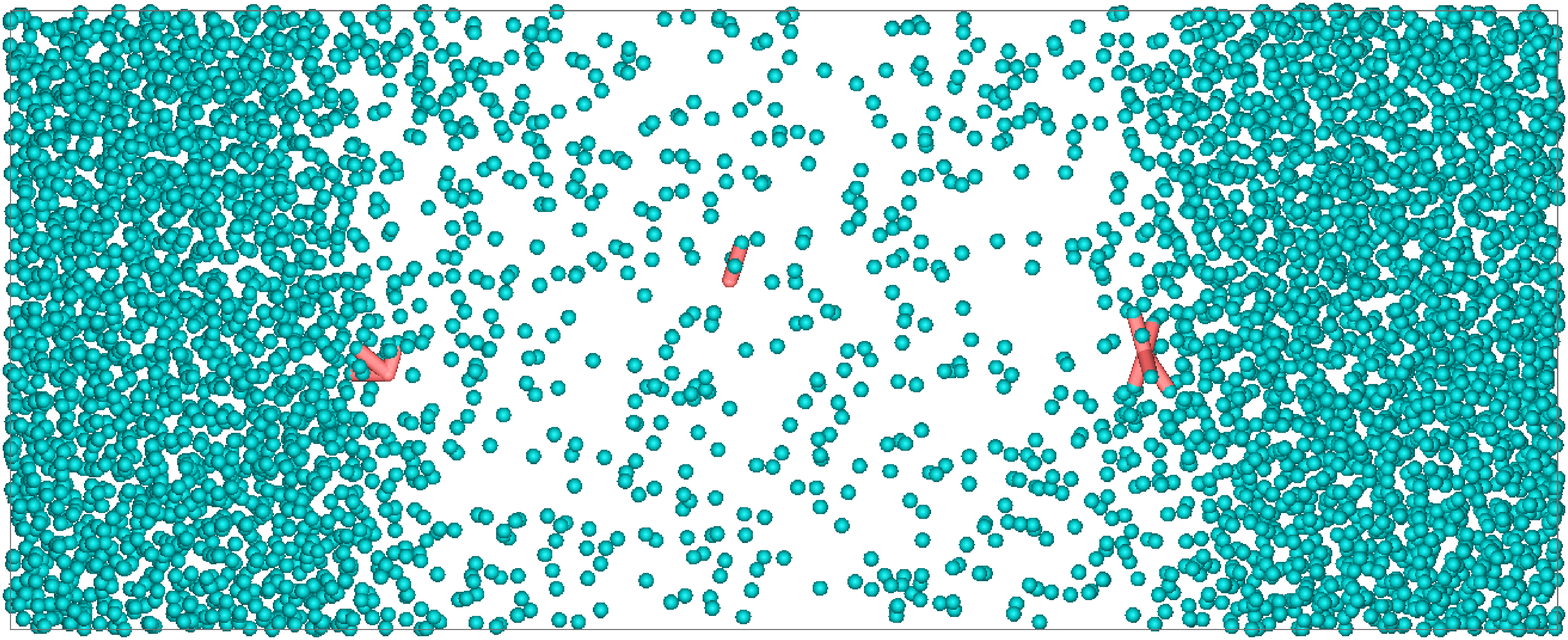}
\includegraphics[width=0.45\textwidth]{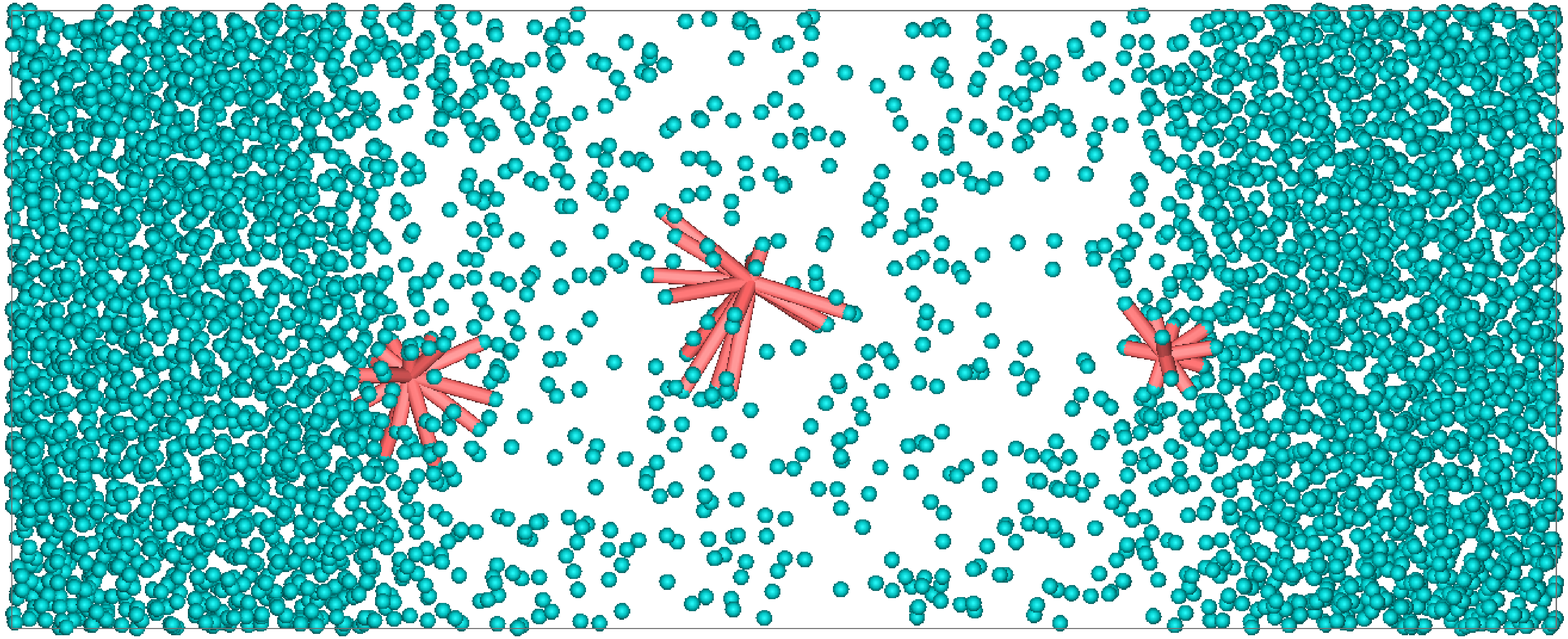}
\includegraphics[width=0.45\textwidth]{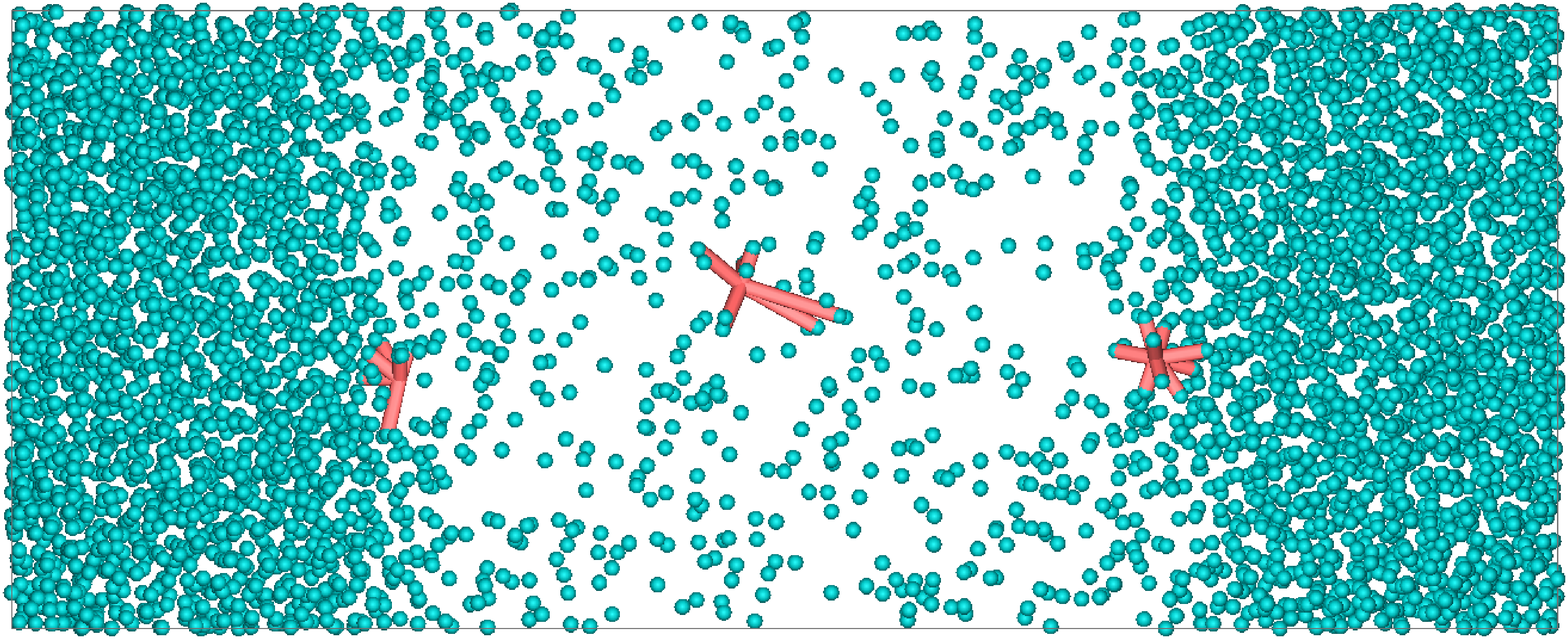}
\caption{
Visual representation of the three algorithms in a two-phase liquid-vapor Lennard-Jones
system: fixed-distance cutoff (top), Voronoi construction (center) and SANN (bottom). In
each case, we select the same particles and check which neighbors are detected using each
algorithm.
}
\label{fig:picture}
\end{figure}

In Figure~\ref{fig:tp} we plot the average number of nearest neighbors $\Big< N_{Nb}(x) \Big>$ and
the corresponding variance $Var(x) = \Big< N_{Nb}^2(x) \Big> - \Big< N_{Nb}(x) \Big>^2$ as a function of $x$ for
both systems. As expected, the liquid-crystal interface (Fig.~\ref{fig:tp} panel a) is barely visible from
all three methods and the results are quite similar. However, while the fixed-distance
cutoff and SANN methods seem to show a slight increase in $\Big< N_{Nb}(x) \Big>$ in the
crystal phase, this does not appear evident in the Voronoi algorithm. In all cases,
the crystal seems to have a slightly lower variance than the fluid. In contrast, the liquid-vapor
interface (Fig.~\ref{fig:tp} panel b) is very well captured by the average number of nearest
neighbors (upper panels) computed by the fixed-distance cutoff, whereas the SANN algorithm shows only a
slight decrease of the number of nearest neighbors in the vapor phase. The Voronoi algorithm finds
even more neighbors in the low-density vapor phase and its standard deviation (lower panel) increases
strongly in the vapor phase. This behavior reflects the strong sensitivity of the Voronoi construction 
to thermal fluctuations. Although $Var(x)$ also fluctuates in the other two methods, 
the changes are much less pronounced than in the Voronoi case.

To get a better understanding for the nearest neighbors found at the interface and in
the vapor phase, in Figure~\ref{fig:picture} we show a snapshot from the two-phase
liquid-vapor system where the vapor phase has been shifted to lie in the center of
the box. We have selected three particles (two at the liquid-vapor interface and one
in the vapor phase) and have calculated their nearest neighbors using all three algorithms.
As expected, in or at the vapor phase the fixed-distance cutoff (upper panel) finds few neighbors
when the cutoff is not tuned to the vapor phase. Note that tuning the fixed-distance cutoff for
the liquid and vapor phases simultaneously is not possible. The Voronoi algorithm (center panel)
detects many neighbors, both for particles at the interface and particularly for particles in
the vapor phase. At the interface the SANN algorithm (lower panel) finds neighbors mostly from
the interface and liquid and, unlike the Voronoi construction, not far-off in the vapor.


\subsection*{Application to bond-order parameters}

\begin{figure}
a)\includegraphics[width=0.45\textwidth]{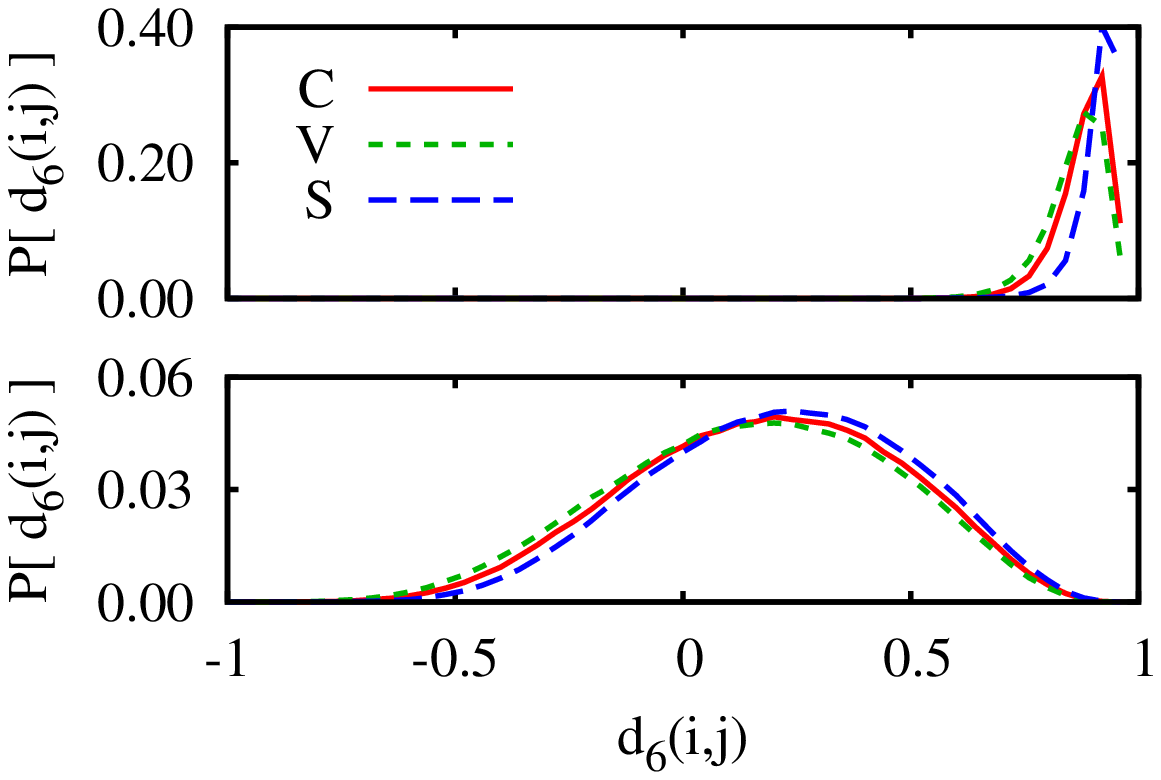}
b)\includegraphics[width=0.45\textwidth]{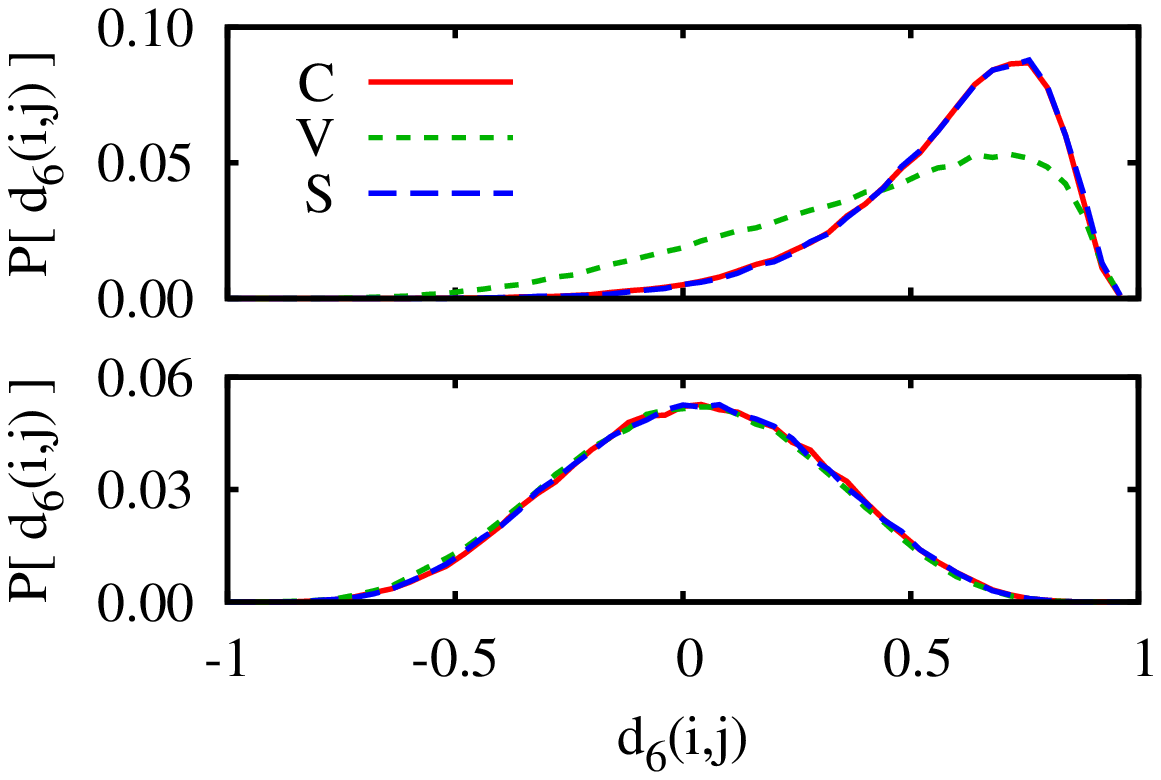}
c)\includegraphics[width=0.45\textwidth]{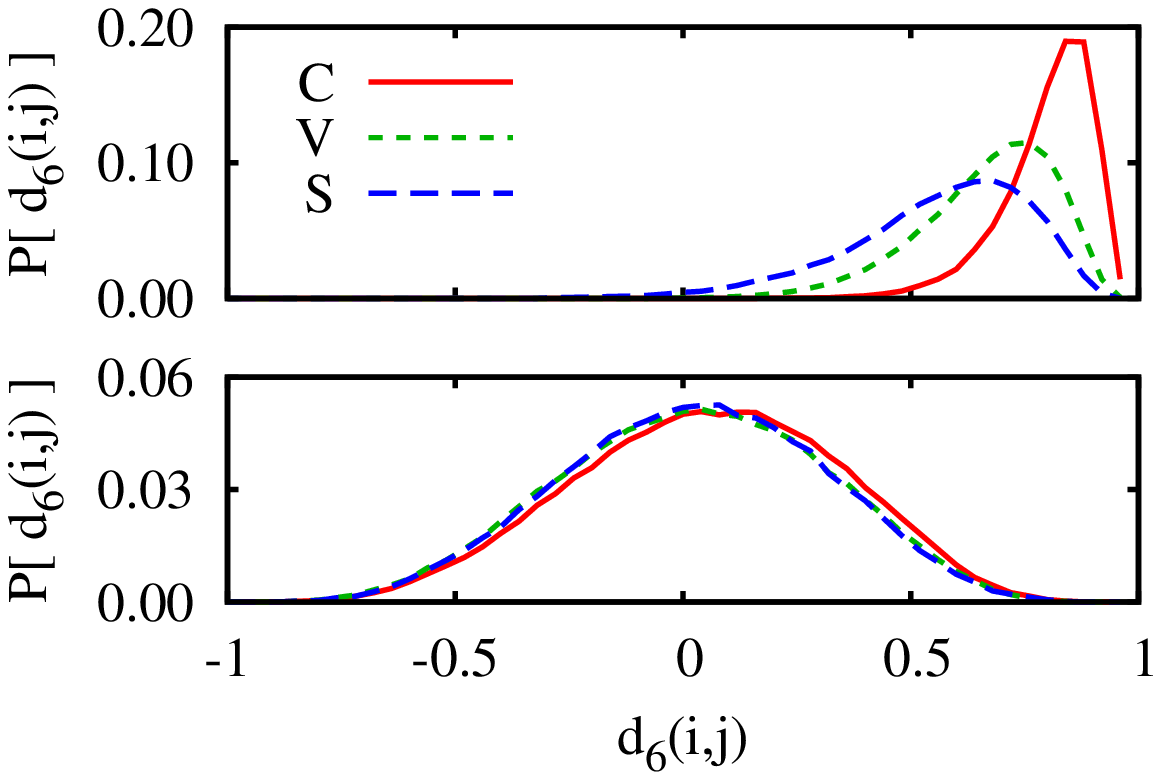}
\caption{Distribution of the local bond-order correlator $d_6(i,j)$ using different
neighbor criteria. Panel a) shows results for Lennard-Jones fcc crystal (upper panel)
and liquid phases (lower panel), panel b) for the $3$-fold coordinated carbon graphite
(upper panel) and liquid (lower panel) phases, and panel c) for the $4$-fold coordinated
carbon diamond (upper panel) and liquid (lower panel) phases. For the carbon phases the
fixed-distance cutoff distance was set to $2.7$ to include the next-nearest neighbors,
as do inherently both the Voronoi and SANN algorithms.
}
\label{fig:bo}
\end{figure}

Finally, we apply the neighborhood algorithms to investigate their effect on the local
bond-order parameters used when studying crystal nucleation. To choose the order of the
spherical harmonics in Eqn.~\ref{eqn:bond}, we match the symmetry of the spherical harmonics, 
i.e. $l$, to the symmetry of the crystal under study. For the
Lennard-Jones system we use $l=6$ due to the close-packed crystalline structure
of the fcc (this is also what is typically used to study hard-sphere systems). 
In the original article on diamond nucleation~\cite{Ghiringhelli_2007}, $l=3$ was applied
to grow both carbon crystal phases since this order parameter is not able to distinguish
between graphite and diamond structures. The symmetry $l=3$ was required since only the
first neighbor shell was taken into account. However, in the present analysis, both
Voronoi and SANN algorithms resulted in neighbor lists which included more than the first
neighbor shell. Consequently, the symmetry of the neighborhood changes, and
$l=6$ becomes perfectly commensurate with the symmetry of this extended environment.
Therefore, we settled for $l=6$ for all systems and set the fixed-distance cutoff in the
carbon case to $2.7$ (the minimum after the second peak of the $g(r)$) to include next-nearest
neighbors. Also, because bond-order correlators are very sensitive to asymmetries in the
nearest neighbor sets and both the fixed-cutoff and the Voronoi construction feature
pair-wise symmetry, we decided to enforce this symmetry for SANN, too, by removing asymmetric
neighbor pairs. Note that the choice to remove neighbors rather than to add them is arbitrary.
The results for the local bond-order correlators distribution $P[ d_6(i,j) ]$ are
presented in Figure~\ref{fig:bo}.

The main criterion for a good order parameter is to have as little overlap as possible
between the crystal phase (upper panels) and the (meta-stable) liquid phase (lower panels).
As shown in Figure~\ref{fig:bo}a all neighborhood algorithms perform reasonably well for
the Lennard-Jones system. However, for the graphite crystal (Fig.~\ref{fig:bo}b upper panel)
both SANN and the fixed-cutoff algorithms allow one to distinguish between a liquid and a solid 
environment, whereas the Voronoi construction shows a severe broadening of the distribution of
the graphite, causing a substantial overlap with the liquid's distribution (lower panel). With such
an overlap the bond-order parameters would fail to distinguish between liquid-like and solid-like
particles. In the diamond case (Fig.~\ref{fig:bo}c) the SANN algorithm performs worse than
the others and again the fixed-distance cutoff works best. In the graphite case the failure of
the Voronoi construction can be attributed to the (arguably) spurious neighboring particles
located in different graphite layers, as previously discussed (Fig.~\ref{fig:graphite-slab}).
The behavior of the SANN algorithm in the diamond case finds its origin in that not enough
neighbors from the second neighbor shell are included (on average a total of $12$ instead of
$16$), and bond-order parameters are particularly sensitive to missing or additional neighboring
particles. 

\subsection*{Benchmarking}

Finally, we measure the run-time of each algorithm for all bulk phases discussed previously.
The benchmark was performed on a computer equipped with an Intel Core2 Quad Q9550 processor
running at $2.83$ GHz and $4$ GB of DDR2 RAM running at $1066$ MHz. All source code was
compiled using the GNU gcc compiler version 4.5.1. The operating system was a $64$-bit
OpenSuSE Linux with Kernel 2.6.37. Each algorithm was running single-threaded and computed
the neighbor sets for all particles in the system. To improve accuracy whenever the run-time
was near our time resolution we measured the total time of $10$ sequential repetitions. All
data presented are averages over at least 3 separate program runs. 

\begin{table}
\begin{tabular}{l||r|r|r||r|r}
System & C & V & S & V/C & S/C \\
\hline
Lennard-Jones liquid & 25 & 610 & 45 & 24.4 & 1.8\\
Lennard-Jones fcc & 20 & 753 & 48 & 37.7 & 2.4 \\
Hard-Spheres $\phi = 0.54$ & 507 & 14390 & 1022 & 28.4 & 2.0\\
Hard-Spheres $\phi = 0.61$ & 528 & 15050 & 1091 & 28.5 & 2.1\\
Carbon $3$-fold liquid & 5 & 160 & 8 & 32.0 & 1.6 \\
Carbon $3$-fold graphite & 5 & 190 & 7 & 38.0 & 1.4\\
Carbon $4$-fold liquid & 6 & 153 & 10 & 25.5 & 1.7\\
Carbon $4$-fold diamond & 6 & 180 & 9 & 30.0 & 1.5
\end{tabular}
\caption{Run-times in milli-seconds of the fixed-distance cutoff (C), the Voronoi construction (V) and
the SANN algorithm (S), and their ratios V/C and S/C. For details on the system samples we refer to
Section~\ref{sec:samples}, for implementation details to Section~\ref{sec:implementation}, and
for the benchmarking procedure to the main text.}
\label{tab:timings}
\end{table}

The timings are presented in Table~\ref{tab:timings}. Compared to the fixed-distance cutoff
the Voronoi construction takes $24.4$ to $38.0$ times longer to compute. In contrast, the
computational cost of SANN is only $1.4$ to $2.4$ times that of the fixed-distance cutoff
and thereby outperforms the Voronoi construction by an order of magnitude. Therefore, we consider
SANN well-suited for application on-the-fly in simulations. 

As a final remark we like to point out that timing results are highly implementation dependent and
as such should be considered as indications only. On the one hand, the CGAL library used for the Voronoi
construction is reasonably fast, but faster implementations may be available. On the other hand, our
implementations of the fixed-distance cutoff and the SANN algorithm may have room for optimization, too.

\section{Conclusion}\label{sec:conclusions}

In this paper we have described an algorithm to compute a particle's nearest neighbors
in an arbitrary  many-particle system. The algorithm is similar to a fixed-distance cutoff
in that all particles within a
cutoff distance are considered nearest neighbors. But rather than using one cutoff
for all particles, this algorithm assigns to each particle an individual cutoff
distance, thereby making it suitable for systems with inhomogeneous densities, such
as gravitational or multi-phase systems. The cutoff distance follows from a geometric
requirement, namely that the sum of all solid angles associated with neighboring
particles adds up to $4\pi$. Thus, the algorithm becomes parameter-free and scale-free.
Though the approach was inspired by the Voronoi construction, it has several advantages
over it: the presented algorithm is significantly easier to implement, computationally
less expensive and more robust against thermal fluctuations.

We tested the algorithm on a number of bulk phases including supercooled liquid and
crystal phases of Lennard-Jones particles, hard spheres and $3$-fold and $4$-fold
coordinated carbon. We compared the nearest-neighbor distributions obtained from SANN
to both the fixed-distance cutoff criterion and the Voronoi construction. In the case
of the Lennard-Jones and hard-sphere phases, our algorithm reproduces very well the
nearest-neighbor distribution of a well-tuned fixed-distance cutoff. This is in
contrast to a Voronoi construction, which has large fluctuations  and
as such does not perform as well. For the carbon phases, our algorithm includes the
second neighbor shell, like the Voronoi construction, but avoids neighbors in the 
neighboring graphite layers.

We also examined particles at the interface of two-phase systems, such as Lennard-Jones
liquid-vapor and liquid-crystal systems. We find that when two high-density phases coexist,
all algorithms give reliable results. However, at the interface between a fluid and a
low-density vapor, our algorithm is more robust  to thermal fluctuations than
the Voronoi construction.

We then employed the neighbor information of all algorithms as input for a bond-order
analysis, which is typically used in crystal nucleation studies for the identification
of solid-like particles in a supercooled metastable liquid. Comparing the bond-order
correlator distributions, we found little difference between the algorithms for the
Lennard-Jones system, indicating that all algorithms are suitable for structure
analysis of close-packed systems. However, the Voronoi construction failed for the graphite
phase due to the identification of spurious neighbors located in different graphite layers,
and SANN performed poorly in the diamond phase, due to the fact that not enough neighbors
from the second neighbor shell were included. Hence, care has to be taken when applying
either one SANN or the Voronoi construction to open structures and network formers. But
where the Voronoi construction fails SANN might succeed, and vice-versa.

Finally, we performed benchmarks on the run-time for all algorithms. On all systems tested
we found the computational cost of SANN to be at most $2.4$ times that of the fixed-distance
cutoff and in all cases it outperformed the Voronoi construction by at least an order of
magnitude.

To conclude, when studying a system at several concentrations or a heterogeneous system,
the proposed algorithm has the advantage that is does not require tuning a parameter for every
concentration/environment. Given the robustness and low computational cost of our algorithm, we argue 
that SANN is well suited not only for post-analysis, but also on-the-fly in simulations. 
It reliably identifies the nearest neighbors, and its behavior for graphite and at a two-phase
interface suggests its application to situations where the Voronoi construction suffers
from distorted polyhedra, like in structural analysis of protein folding trajectories~\cite{Poupon_2004,Cazals_2006},
in DNA-mediated colloidal crystallization~\cite{Jahn_2010,Geerts_2010}, in suspensions of patchy colloids 
with tetrahedral or octahedral symmetry~\cite{Voivod_2011,Noya_2010,Noya_2007,Romano_2011} and in
water~\cite{Abascal_2005}. Finally, the SANN algorithm is not only useful for 
simulation data, but should also be useful in analyzing experimental 3D images, as obtained, for instance, 
by confocal microscopy or by tomography.

\begin{acknowledgments}
The authors thank K. Shundyak for fruitful discussions. The work of JvM at the FOM Institute
is part of the research program of FOM and is made possible by financial support from
the Netherlands Organization for Scientific Research (NWO). DF acknowledges financial
support from the  Royal Society of London (Wolfson Merit Award) and from the ERC
(Advanced Grant agreement 227758). CV acknowledges support from an Individual Marie Curie
Fellowship (in Edinburgh) and from a Juan de la Cierva Fellowship (in Madrid). LF
acknowledges support from the EPSRC, U.K. for funding (Programme Grant EP/I001352/1). 
\end{acknowledgments}

\end{document}